\documentclass[universe,review,accept,pdftex,moreauthors]{Definitions/mdpi} 

\usepackage[labelformat=simple]{subcaption}

\DeclareCaptionLabelFormat{subcaptionlabel}{\normalfont(\textbf{#2}\normalfont)}
\usepackage{breakurl}
\firstpage{1} 
\pubvolume{10}
\issuenum{2}
\articlenumber{67}
\pubyear{2024}
\copyrightyear{2024}
\externaleditor{Academic Editor: Benjamin Grinstein and Lijing Shao}
\datereceived{22 November 2023} 
\daterevised{31 December 2023} % Comment out if no revised date
\dateaccepted{17 January 2024} 
\datepublished{1 February 2024} 
\hreflink{https://doi.org/10.3390/\linebreak~universe10020067} % If needed use \linebreak
\usepackage{wasysym}
\usepackage{array, multirow, bigdelim, makecell, booktabs} 
\usepackage{morefloats}
\extrafloats{100}
\usepackage{comment}
\usepackage{bm}% bold math
\usepackage{slashed}
\usepackage[normalem]{ulem}
\usepackage{pifont}
\usepackage{color}
\usepackage{array}

\usepackage{makecell}
\definecolor{nicered}{rgb}{0.7,0.1,0.1}
\definecolor{nicegreen}{rgb}{0.1,0.5,0.1}
\Title{Probing Dark Sectors with Neutron Stars}

\TitleCitation{Probing Dark Sectors with Neutron Stars}

\Author{Susan Gardner *$^{,\dagger}$\orcidB{} and Mohammadreza Zakeri $^{\dagger}$\orcidC{}}

\AuthorNames{Susan Gardner and Mohammadreza Zakeri}

\AuthorCitation{Gardner, S.; Zakeri, M.}
\address[1]{%
Department of Physics and Astronomy, University of Kentucky, Lexington, KY 40506-0055, USA; 
 m.zakeri@uky.edu
}
\corres{\hangafter=1 \hangindent=1.05em \hspace{-0.82em}Correspondence: susan.gardner@uky.edu}

\firstnote{\hangafter=1 \hangindent=1.05em \hspace{-0.82em}These authors contributed equally to this work.}

\abstract{Tensions in the measurements of neutron and kaon weak decays, such as of the neutron lifetime, may speak to the existence of new particles and dynamics not present in the Standard Model (SM). In scenarios with dark sectors, particles that couple feebly to those of the SM appear. We offer a focused overview of such possibilities and describe how the observations of neutron stars, which probe either their structure or dynamics, limit them. In realizing these constraints, we highlight how the assessment of particle processes within dense baryonic matter impacts the emerging picture---and we emphasize both the flavor structure of the constraints and their broader connections to cogenesis models of dark matter
and baryogenesis.}

\keyword{neutron stars; baryon number; neutron lifetime anomaly; dark matter; dark sectors} 

\begin{document}
%%%%%%%%%%%%%%%%%%%%%%%%%%%%%%%%%%%%%%%%%%

% ========================================================
\section{Introduction}
\label{sec:introduction}
% ========================================================
Various astronomical observations speak to the existence of dark matter and dark energy~\cite{Zwicky:1933,Zwicky:1937, peebles1970adiabatic,Press_Schechter:1974,1978MNRAS.183..341W,1982ApJ...263L...1P,blumenthal1984colddarkmatter,Davis_Frenk_White:1985,Rubin_Ford:1978,Faber_Gallagher:1979,Rubin_Ford:1980,hu2002reviewcmb,Planck2020legacy,Eisenstein_SDSS:2005ApJ...633..560E,Reiss:1998AJ,Perlmutter_1999}, and~the detections of neutrinos from extraterrestrial sources reveal nonzero neutrino masses~\cite{Davis:1968cp,Super-Kamiokande:1998kpq,SNO:2002tuh}. Also, the~cosmic excess of matter over antimatter~\cite{Sakharov:1967dj,Planck:2018vyg}, i.e.,~the baryon asymmetry of the universe (BAU), cannot be described within the SM, because~the electroweak phase transition is not of the first order~\cite{Kajantie:1996mn,Csikor:1998eu} for the empirically determined Higgs mass~\cite{ATLAS:2012yve,CMS:2012qbp}. Thus, physics beyond the SM has been established, even if its origins continue to be mysterious. Although~these may come from dynamics that is disjoint from the SM, the~possibility of identifying its nature mandates the careful consideration of any experimental anomalies that appear, for~they could point to new dynamics. In~this article, we consider the neutron lifetime anomaly~\cite{Serebrov:2008her,Wietfeldt2011RvMP...83.1173W}, which entails the~possibility that it could, at~least in part, derive from new physics~\cite{Fornal2018PhRvL.120s1801F,Berezhiani:2018eds,Barducci:2018rlx,Rajendran:2020tmw}, and~how that possibility can be constrained by observations of neutron stars~\cite{Nelson:2018xtr,Baym2018PhRvL.121f1801B,Motta:2018rxp,Berezhiani:2020zck,Berryman:2022zic,Berryman:2023rmh}. To~set the stage, in~this introduction we (i) place the neutron lifetime anomaly in the context of other low-energy $\beta$ decay experiments, (ii) motivate why new physics in neutron decay could be linked to models that explain both dark matter and the BAU, and~(iii) discuss how neutron dark decays connect to neutron star structure and~dynamics.

\subsection{Low-Energy Experimental Anomalies in Beta~Decay}

The expected pattern of quark flavor mixing under charged current interactions in the SM make the Cabibbo angle, i.e.,~the mixing of the first- and second-generation of quarks, central to a unitarity test of the Cabibbo--Kobayashi--Maskawa (CKM) matrix with elements $V_{ij}$. This test probes the possibility of additional contributions to quark mixing and/or whether $G_F$, the~effective strength of the weak interactions as measured through the muon lifetime, is really universal. The~so-called Cabibbo angle anomaly (CAA) has different components, and~its resolution gains in importance, because possible non-SM effects need not be very small~\cite{Coutinho:2019aiy,Crivellin:2023zui}. There are disagreements in the assessment of $V_{ud}$, which is most precisely probed through neutron decays or through superallowed ($0^+ \to 0^+$) transitions in nuclei, as well as in $V_{us}$, for~which $K \to \ell \nu_{\ell}$ ($K_{\ell2}$) and $K \to \ell \nu_{\ell} \pi$ ($K_{\ell3}$) decay studies are the most precise. Theoretical corrections and inputs are needed in the analysis of the experiments to determine $V_{ij}$. The~determinations of $V_{us}$ from $K_{\mu 2}$ and $K_{\ell 3}$ decays do not agree~\cite{Moulson:2017ive,Seng:2021nar,Cirigliano:2022yyo}, and~the extracted values of $V_{ud}$ are also in slight tension with each other~\cite{Cirigliano:2022yyo}. The~combined results are also at odds with CKM unitarity, with~the differences predicated by the computed radiative corrections~\cite{Czarnecki:2004cw,Marciano:2005ec,Seng:2018qru}. In~the superallowed case, the~effect of the axial vector current enters in the radiative corrections, which can be modified according to nuclear structure as well~\cite{Towner:2002rg}. Recent developments in the theory of the radiative corrections in neutron $\beta$ decay have led to refined assessments of their value, thus impacting the value of $V_{ud}$~\cite{Seng:2018qru}; we refer to~\cite{Universe:GorchSeng} in this volume for a summary.  It is worth noting that these determinations also differ from the value of $|V_{us}/V_{ud}|$ determined from the ratio of rates in $K_{\ell \nu}$ and $\pi \to \ell \nu_{\ell}$ ($\pi_{\ell\nu}$) decays~\cite{Seng:2021nar,Cirigliano:2022yyo}. These anomalies could have various resolutions, thereby arising from either limitations in the experiments themselves (there is only one high quality measurement of $K\to \mu 2$) or in the SM theoretical inputs; however,~particular new physics models, such as those with right-handed currents, could be at work~\cite{Cirigliano:2022yyo,Crivellin:2023zui}.

The mixed symmetry nature of neutron $\beta$ decay implies that at least two different experiments are needed to extract $V_{ud}$, and~our interests make it important to flesh out the details. In~the SM, the~neutron lifetime  $\tau_n$ is

\vspace{-6pt}
\begin{equation}
    |V_{ud}|^2 \tau_n ( 1 + 3 \lambda^2) = \frac{2\pi^3}{G_F^2 m_e^5 (1+ \delta_{\rm RC})f} \equiv \eta \,,
    \label{taun_tie}
\end{equation}
where $\lambda\equiv g_{A}/g_{V}$ (the weak coupling constants of the nucleon), $m_e$ is the electron mass, $f$ is a phase space factor,  and~$\eta$ is $4908.6 (1.9)\,{\rm s}$~\cite{Marciano:2005ec},~$4903.6 (1.0)\,{\rm s}$~\cite{Seng:2018qru}, or $4905.7 (1.5)\,{\rm s}$~\cite{Czarnecki:2019mwq}, depending on the calculation of the radiative correction $\delta_{\rm RC}$ (and $f$) used. Following common practice, we use  $\lambda$ and $g_A$ interchangeably, which is also strongly supported by an analysis of the radiative corrections to $g_A$~\cite{Gorchtein:2021fce}. The~uncertainties in $\delta_{\rm RC}$ dominate those in $\eta$ and have been used to determine its reported error. Thus, we see that the right-hand side of Equation~(\ref{taun_tie}) is strongly constrained, and~the parameter $\lambda$ is determined separately through measurements of the neutron $\beta$ decay correlations, with~the coefficients~\cite{Jackson:1957zz} termed as the ``neutron alphabet~\cite{Abele:2008zz}''. Thus, to determine $V_{ud}$, in order to~combine with $V_{us}$ for a test of CKM unitarity, both the lifetime and a measured decay correlation are needed. The~most precisely determined correlation coefficient is ``$A$'', which is the correlation between the neutron spin and the momentum of the decay electron;~thus, that determination has been commonly used. Recently, precise determinations of ``$a$'', which is the correlation of the decay electron and antineutrino momenta, have been achieved. Thus, we shall compare the outcomes of combining the measurements of either the $A$ or $a$ correlation coefficients with the neutron lifetime for $V_{ud}$. 

\begin{figure}[H]
 
\includegraphics[width=10.5 cm]{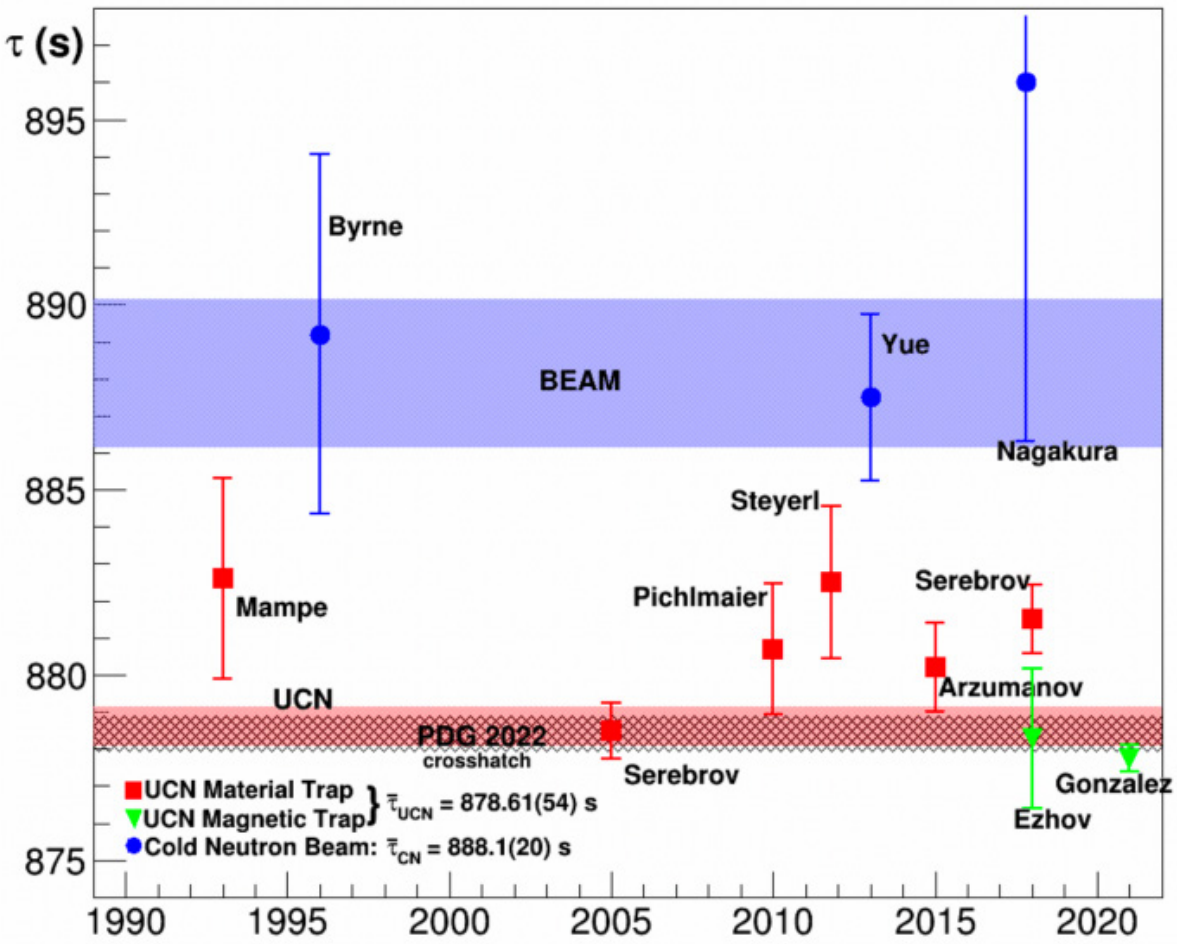}
\caption{\label{fig:tau_n} Measured values of the neutron lifetime from beam (blue), material trap (red), and magnetic trap (green) experiments since 1990 from~\cite{LRP_2023}. In~all neutron trap experiments, which use ultracold neutrons (UCNs), the~neutron lifetime was determined by counting the neutrons that persist, whereas in the beam experiments, it was determined by counting the decay protons. The~central values of the two sets of experiments disagree by $9.5\,\rm s$, with~a significance in excess of $4\,\sigma$---this is the neutron lifetime anomaly. The~Particle Data Group (PDG) does not include the beam measurements in their fits~\cite{ParticleDataGroup:2022pth}. }
\end{figure}

In Figure~\ref{fig:tau_n}, we show the data set that constitutes the neutron lifetime anomaly---we see that the measured lifetime $\tau_n$ has been shown to be significantly different if the neutrons decay while stored in a bottle or~trap, with~the surviving neutrons counted after some storage time, or~if the neutrons decay while in a beam within a device that counts the decay protons. The~determined neutron lifetimes differ by some $9.5\, \rm s$, which is a~significant variation. One prominent impact this shift could have is on big bang nucleosynthesis (BBN), because~the neutron lifetime determines, to~a good approximation, the~temperature at which the weak interactions freeze out in the early universe---and a shorter lifetime reduces the expected yield of $^{4}{\rm He}$~\cite{Mathews:2004kc}. We refer to~\cite{Universe:BBN_Yeh} for a review of the neutron lifetime's impact on element yields from BBN. Since the bottle lifetime is shorter, the~difference between the two determinations can be reasonably attributed to the possibility of neutron dark decays~\cite{Fornal2018PhRvL.120s1801F} coming from physics beyond the SM, and~we refer to the review~\cite{Universe:Fornal} in this volume for further~discussion. 

% =======================================================

% =======================================================

\begin{figure}[H]
 
\includegraphics[width=\textwidth]{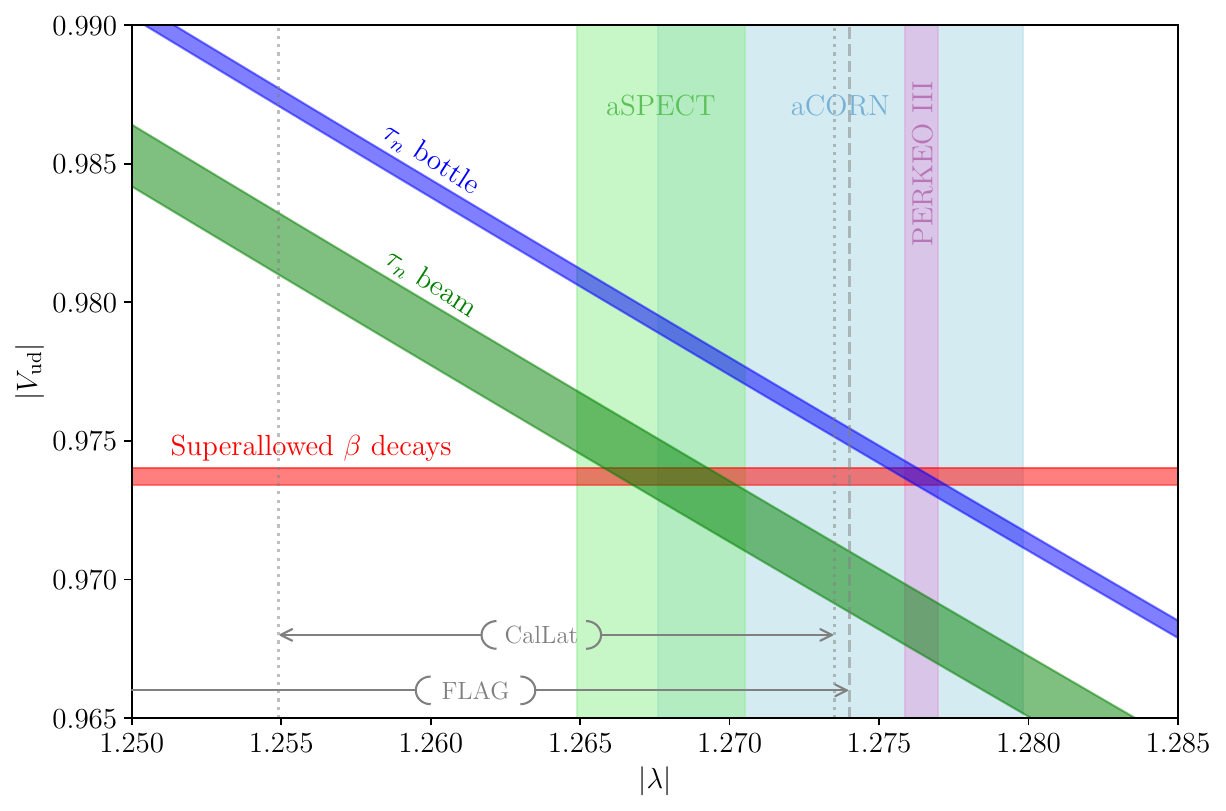}
    \caption{\label{fig:decaycorr}
    The SM correlation between the CKM matrix element \( |V_{ud}| \) and the axial vector to vector coefficient ratio \( |\lambda| = |g_A / g_V| \) is presented utilizing the averaged neutron lifetime values from bottle (blue) and beam (green) methods given in Figure~\ref{fig:tau_n}. This figure includes radiative corrections as per~\cite{Seng:2018qru}. Vertical bands illustrate \( |\lambda| \) measurements from aSPECT~\cite{Beck:2019xye} (light green), aCORN~\cite{Wietfeldt:2023mdb} (light blue), and~PERKEO III~\cite{Markisch:2018ndu} (light purple). The~measured value of \( V_{ud} \) via superallowed nuclear transitions is shown by a red band~\cite{ParticleDataGroup:2022pth}. Additionally, lattice values of \( g_A \) are shown comprising the 2021 FLAG average for \( N_f =2+1+1 \) flavors~\cite{FlavourLatticeAveragingGroupFLAG:2021npn} and the 2019 results from the CalLat collaboration~\cite{Walker-Loud:2019cif}.
    }
\end{figure}

Testing the $V-A$ structure of neutron (and nuclear) $\beta$ decay in the SM has been a focus of intense research for decades~\cite{Abele:2008zz, Dubbers:2011ns, Cirigliano:2013xha, Dubbers:2021wqv}. Here, the various decay observables, including the lifetimes and $\beta$ decay correlations, can constrain the possibility that the neutron lifetime anomaly comes from new physics~\cite{Czarnecki:2018okw,Dubbers:2018kgh,Berryman:2022zic}. That is, the~agreement of the value of $V_{ud}$ with CKM unitarity and the value of $\lambda$ in a neutron lifetime and in a decay correlation limit the possibility of non-$V-A$ contributions, which could include neutron dark decays or the possibility of $n-n'$ oscillations, to~the lifetime. In~that context, the~PDG gives an average of $|\lambda|=1.2754(13)$ with a scale factor of 2.7 from the $A$ and $a$ correlation coefficients~\cite{ParticleDataGroup:2022pth}, where the most precise determination, from~an $A$ measurement, is from PERKEO III $|\lambda|=1.27641(56)$~\cite{Markisch:2018ndu}. The~other two most precise determinations of $\lambda$, which are also determined from $A$~\cite{Mund:2012fq,UCNA:2017obv}, are consistent with that result. Recent and precise $a$ determinations~\cite{Beck:2019xye,Beck:2023hnt,Wietfeldt:2023mdb} have emerged, however, that differ significantly from the $A$ results. We note, e.g.,~$|\lambda|=1.2677(28)$~\cite{Beck:2019xye}---and the latest status of the various measurements is shown in Figure~\ref{fig:decaycorr}. This discrepancy cannot yet be challenged by direct calculation in lattice quantum chromodynamics (QCD). As~for the lattice QCD results, we note the Flavour Lattice Averaging Group (FLAG) average from simulations with $N_f = 2 + 1 +1$ flavors, $g_A=1.246 (28)$~\cite{FlavourLatticeAveragingGroupFLAG:2021npn}, as~well as the CalLat-18 result of $g_A=1.271 (13)$~\cite{Chang:2018uxx}, which is compatible with the PERKEO III determination, though~it has been updated to $g_A=1.2642 (93)$~\cite{Walker-Loud:2019cif}. Placed in the context of a test of CKM unitarity, as~shown in Figure~\ref{fig:decaycorr} using the radiative corrections of~\cite{Seng:2018qru}, we see that the values of $|V_{ud}|$ from the bottle lifetime and the PERKEO III $A$ measurement are entirely consistent with the value determined in superallowed decays. Interestingly, the~value of $|V_{ud}|$ from the measured beam lifetime combined with $\lambda$ from an $a$ measurement is very comparable to its value determined using the bottle lifetime and the $\lambda$ from $A$. In~the presence of two possible solutions for $\lambda$ and $V_{ud}$ that are similarly compatible with CKM unitarity, it would seem we cannot (yet) limit a dark decay solution to the neutron lifetime anomaly on the basis of this comparison. This perspective differs from that of earlier work~\cite{Dubbers:2018kgh,Czarnecki:2018okw}, and~it is driven by the recent $a$ measurements. Ultimately, any solution to the neutron lifetime anomaly using new physics is required to be compatible with the measured $V-A$ structure of the SM~currents. 

% =======================================================
%              (lambda, V_ud) Figure
% =======================================================
\unskip
% =======================================================

\subsection{From Anomalies to Dark Matter and the Cosmic Baryon~Asymmetry}
\begin{figure}[H]
\includegraphics[width=0.8\textwidth]{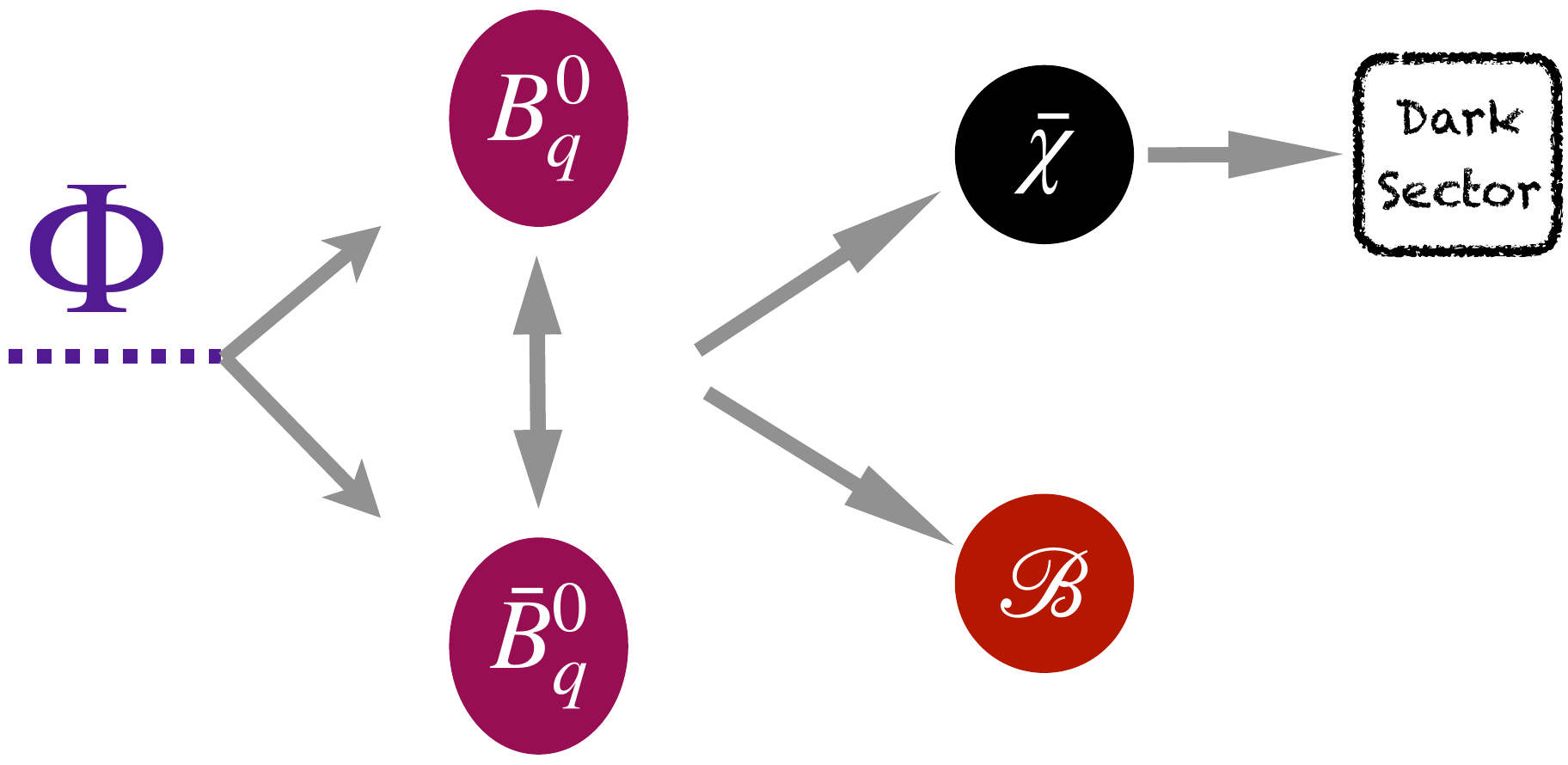}
    \caption{\label{fig:meso_schem}
    A schematic illustration of $B$ mesogenesis~\cite{Elor:2018twp}: a~late-scale dark cogenesis model. Here, a scalar $\Phi$, produced out of equilibrium, decays and hadronizes to a $B_q^0 {\bar B}_q^0$ pair, with~$q\in d,s$. The~pair mixes in~the presence of CP violation, and, finally, a~$B$ meson decays preferentially to ${B} \to {{\cal B}} \bar\chi$,  which is a~baryon ${\cal B}$---a dark antibaryon ${\bar \chi}$ pair. Since ${\bar \chi}$ carries ${\cal B}=-1$, the ${\cal B} \bar{\chi}$ final states do not break the baryon number, yet this scenario produces a baryon asymmetry in the visible sector. The~produced ${\bar \chi} (\chi)$ decays to other dark sector states, thus preventing washout of the baryon asymmetry, with~the particles of the dark sector final states serving as dark matter candidates. Different dark sector choices, that we note in text, differ in their impacts on the dynamics of a neutron star~\cite{Berryman:2023rmh}.
     }
\end{figure}

Although astronomical observations speak to the existence of dark matter that is (i) stable or nearly so on Gyr time scales~\cite{Planck2020legacy,Eisenstein_SDSS:2005ApJ...633..560E}, (ii) possessive of no substantial strong or electromagnetic charge~\cite{Clowe:2006eq}, and~(iii) engenders an ``inside-out'' formation of large-scale structure~\cite{blumenthal1984colddarkmatter}, many different sorts of dark matter candidates are possible~\cite{Battaglieri:2017aum,Chou:2022luk}, and~they can impact astrophysical observations at small scales differently~\cite{Gardner:2021ntg}. It is intriguing that the contributions of baryonic matter and of dark matter to the present cosmic density parameter $\Omega_0$ are not too dissimilar, with~the fraction in baryonic matter being some 20\% of the dark matter one~\cite{Planck2020legacy}. Perhaps, then, dark matter is asymmetric~\cite{Nussinov:1985xr,Barr:1990ca}---much as baryonic matter is---thereby suggesting that the two effects could share a common origin. That possibility could be realized through a dark cogenesis model~\cite{Davoudiasl:2012uw,Elor:2018twp,Alonso-Alvarez:2021oaj}, and~interestingly
%!Author: fixed typo
%m 
there is also a connection between such models and a dark decay model of the neutron lifetime anomaly~\cite{Fornal2018PhRvL.120s1801F}, thereby yielding the dark decay $n\to \chi\gamma$, where $\chi$ is a fermion with a baryon number of ${\cal B}=1$. This connection is concrete, even if the ultimate numerical impact of the $n\to \chi \gamma$ decay rate on the neutron lifetime anomaly finally turns out to be small. Dark cogenesis models can act in counterpoint to other popular mechanisms, such as via weak-scale supersymmetry---in the latter, the severity of permanent EDM searches for the electron and neutron limit the possible phase space severely~\cite{Morrissey:2012db}. In~contrast, the~$B$ mesogenesis~\cite{Elor:2018twp} mechanism and its variants~\cite{Nelson:2018xtr,Elahi:2020urr,Elor:2020tkc,Alonso-Alvarez:2021qfd,Alonso-Alvarez:2021oaj,Elahi:2021jia} do not require sources of CP violation beyond the SM to operate, and~they are, moreover,~testable mechanisms of baryogenesis~\cite{Barrow:2019viz}. %Please check meaning retained 
The schematic illustration of $B$ mesogenesis is presented in Figure~{\ref{fig:meso_schem}}. The~branching ratio for $B\to {\cal B} \bar{\chi}$ decay cannot be too small~\cite{Elor:2022jxy}, and~experimental constraints exist from searches using BaBar data, with~improvements expected from Belle-2~\cite{BaBar:2023rer}. To~generate the noted decays, new scalars ought to exist as well, and~they can be probed through collider searches and through constraints from flavor physics~\cite{Fajfer:2020tqf,Alonso-Alvarez:2021oaj}. Here, we discuss what further constraints come from precision studies of neutron stars and~their dynamics~\cite{Berryman:2022zic,Berryman:2023rmh}. 

% =======================================================
\unskip
% =======================================================

\subsection{Constraints from Neutron Star Structure and~Dynamics}
\label{sec:intro:NSconstraints}

Taken at face value, the~difference in the beam and bottle lifetimes reflects roughly a $1\%$ change in the neutron decay rate---and if the neutron does decay via $n\to \chi\gamma$ or $n\to \chi \phi$, where $\phi$ is a scalar that could also decay via $\phi\to e^+e^-$~\cite{Fornal2018PhRvL.120s1801F}, then the ability to form a protoneutron star is impacted~\cite{Mckeen2018PhRvL.121f1802M,Baym2018PhRvL.121f1801B,Motta:2018rxp}. In~particular, its maximum mass is only some $0.7M_\odot$, which is in~conflict with observations of neutron stars of much larger masses. This constraint can be sidestepped in different ways, such as if $\chi$ self-interactions are included in the model, as~also studied in~\cite{Cline:2018ami} as~an example. We note~\cite{Universe:NS_Husain} in this volume for a discussion of neutron star constraints on dark scalar boson production in this context. Here, rather, we focus on the new physics channels with dark fermions, particularly with a baryon-number-carrying particle $\chi$ and~the connections of this possibility to models that describe dark matter and baryon asymmetry. In~this context, $\chi$ must decay, though~not to antibaryons, to~avoid a washout of the baryon asymmetry, and~the size of the realized baryon asymmetry relies on the rate of $B \to {\bar \chi} {\cal B}$. Here then, the~neutron dark decay rate can be smaller and~thus be in accord with neutron star observations. It thus would not explain the full size of the neutron lifetime anomaly, but~constraints on its size are useful in that they can be interpreted in terms of limits on the flavor structure of the quark--dark baryon couplings. Here we note that although $\chi$ must decay to other dark particles, different dark sector choices can be made. The~precise energy loss constraints we are able to make from the study of existing neutron stars are sensitive to those choices, and~we will explain how in Section~\ref{sec:NS:dark_sectors}. There are also dark decay models, such as the $n\to 3\chi$ model of~\cite{Strumia:2021ybk}, that we cannot constrain with the methods we review in this paper, though~that model is constrained by the would-be agreement between the decay correlations and the bottle lifetime, should the anomaly in the $a$ and $A$ determinations of $\lambda$ (or $g_A$) be~resolved.

In what follows, we discuss, in~turn, the~effects of the existence of dark sectors on stars and, particularly, how neutron stars can constrain models of neutron dark decays, how computations of particle processes in dense matter are important to determining those constraints, and,~finally, what constraints emerge from neutron stars. We then conclude with a perspective on the neutron lifetime anomaly and~a final~summary. 

% ========================================================
\section{Stellar Constraints on Dark~Sectors}
\label{sec:NS:dark_sectors}
% ========================================================
Neutron stars, which are remnants of stellar evolution, present a unique opportunity to study the intricate interactions of dark matter within extreme astrophysical environments. In~this section, we place this in a broader context, 
thus delving into the multifaceted role of dark matter across different stages of stellar evolution, from~the early universe's first stars to the dynamic phenomena observed in supernovae. We explore how dark matter not only influences the life cycle of stars but also plays a crucial role in the formation, structure, and~evolution of neutron~stars.

% ==========================================
\subsection{Dark Matter's Role in Stellar~Evolution}
% ==========================================
The role of dark matter extends beyond the confines of neutron stars. Within~the so-called $\Lambda$ CDM cosmological framework, cold dark matter (CDM) is a key driver of the universe's ``inside-out'' structural growth, thereby crucially shaping the formation of galaxies and stars from the earliest density fluctuations~\cite{1982ApJ...263L...1P}. Its influence is not limited to gravitational interactions: the dark sector potentially engages in a variety of interactions and phenomena that have a lasting impact on stellar evolution. This section explores these interactions, starting with the pivotal role of dark matter in the life cycle of the first~stars.

% ==========================================
\subsubsection{Impact on First Stars and Population I~Stars}
% ==========================================
In the earliest chapters of cosmic history, population III (Pop III) stars, the~universe's very first luminous bodies, emerged from the gravitational collapse of pockets of gas in small halos driven by dark matter~\cite{Ciardi:2004ru}. These stars have been predicted to form at redshifts of $z <$ 20--30 in~halos with masses ranging from $10^{6}\, M_{\odot}$ to $10^{8}\, M_{\odot}$. Recent reports suggest the possible identification of a Pop III star cluster in the remote fringes of a galaxy, thus highlighting ongoing efforts to deepen our understanding of these primordial objects~\cite{Maiolino:2023wwm}. Beyond~its gravitational influence, dark matter, through its annihilation channels, might have significantly impacted the formation of Pop III stars. Such annihilation events could act as an additional heat sources, thereby potentially inhibiting protostar formation by offsetting essential cooling mechanisms~\cite{Spolyar:2007qv, Iocco:2008rb}. In~scenarios where a high dark matter density exists at the gas cloud's location, if~some annihilation products are trapped and thermalized within the cloud, and with this dark matter-induced heating surpassing expected cooling effects, a~novel stellar phase, known as the ``dark star'' phase, could emerge. This phase could manifest itself as a giant hydrogen--helium star ($R \gtrsim 1$ AU), which would be powered by dark matter annihilation instead of nuclear fusion~\cite{Spolyar:2007qv}. Furthermore, with~the ongoing capture of dark matter, these objects may evolve into supermassive dark stars, with~masses larger than $(10^5$--$10^7) M_{\odot}$ and luminosities exceeding $(10^{9}$--$10^{11}) L_{\odot}$~\cite{Freese_2010}. Such objects could potentially be observable by the James Webb Space Telescope (JWST)~\cite{Freese_2010,Zackrisson_2010}. Among~the four metal-poor, high-redshift ($10.3<z<13.2$) sources discovered by the JWST Advanced Deep Extragalactic Survey (JADES)~\cite{Robertson:2022gdk, curtislake2023spectroscopic}, three have been identified as potential supermassive dark star candidates~\cite{Ilie:2023zfv}.

Dark matter can significantly influence the evolution of Pop III stars postformation, especially for those near the Eddington limit. The~additional heating from dark matter annihilation could destabilize the most massive Pop III stars, thereby effectively setting an upper mass limit. For~instance, the~impact of nonthermal, super heavy dark matter particles (mass range $10^8$--$10^{15}$ GeV) is explored in~\cite{Ilie:2019sjk}. It has been found that at a dark matter density of about $10^{16}\, {\rm GeV}/{\rm cm}^3$, conforming to the XENON1T one-year detection bounds~\cite{PhysRevLett.121.111302}, the~maximum mass of a Pop III star could be restricted to a few tens of solar masses. In~environments with extreme dark matter densities, such as $10^{18}\, {\rm GeV}/{\rm cm}^3$, the~growth of Pop III stars might be limited to just a few solar masses due to the Eddington~limit.

Pop III stars, which are luminous yet transient features of the early universe, concluded their short lifecycles in supernovae, thereby enriching the cosmos with heavy elements. However, the~influence of dark matter could extend beyond the inception of these stars, thus significantly shaping the endpoint of their evolution. This influence is epitomized by the black hole mass gap (BHMG), which is a~range delineated by $45\, M_{\odot} \leq M_{\rm BH} \leq 120\, M_{\odot}$ where no black holes emerge from the direct collapse of massive stars due to pair~instability.

The upper boundary of the BHMG results from supermassive stars bypassing pair-instability supernovae to form black holes directly, while the lower threshold arises from less massive stars that undergo pulsational pair-instability supernovae, thereby shedding significant mass through pulsations and leading to lower-mass black hole remnants than their progenitors. Modifications to Pop III stellar physics, potentially through new physics such as light particle emission, can shift the BHMG boundaries. Such particles can expedite helium burning and diminish stellar wind mass loss, thereby consequently reducing the available oxygen for burning and suppressing the pulsational pair-instability mechanisms in Pop III progenitors~\cite{Croon:2020ehi}.
% ==========================================

Transitioning from dark matter considerations in the stellar evolution in the early universe, we now turn our focus to the influence of dark sector physics on more contemporary stars.
Luminosity measurements from the Sun and horizontal branch (HB) stars are instrumental in constraining dark sector particles, which are known for their potential to alter stellar cooling rates. The~Sun's emission of energy through a ``dark channel'', such as dark photons, is particularly restricted by solar boron (${}^8$B) neutrino flux measurements~\cite{Haxton:2012wfz}. These observations establish a conservative upper limit on the Sun's dark luminosity, which is mandated to be less than 10\% of the Sun's total photon luminosity~\cite{Redondo:2013lna}. Furthermore, the~emission of light particles such as neutrinos and axions from the cores of red giants can delay helium ignition. This results in the development of more massive cores, thus leading to a brighter phase prior to their evolution into HB stars~\cite{raffelt1996stars}. The~tip of the red giant branch (TRGB) is especially sensitive to new 
energy loss mechanisms, including those prompted by axions~\cite{PhysRevD.102.083007}, dark photons~\cite{An:2013yfc}, or~millicharged particles~\cite{PhysRevD.43.2314}, which can cause noticeable deviations in TRGB brightness from their standard values. Conversely, dark matter annihilation within a star's core might trigger an earlier helium ignition, which could potentially lead to a dimming of the TRGB~\cite{Lopes:2021jcy}, thereby offering a contrasting perspective on the impact of dark matter within stellar~environments.

Significant dark matter accumulation within stars can impact their structures by altering convective and radiative regions due to energy transfer by dark matter particles~\cite{PhysRevD.95.023507}. These structural changes influence stellar pulsations, with~implications for asteroseismology~\cite{Rato:2021tfc, 2022FrASS...9.8502A}. In~asymmetric dark matter (ADM) scenarios, where a primordial asymmetry determines the dark matter relic abundance instead of annihilation processes~\cite{ZUREK201491}, ADM can accumulate without annihilation counterbalance. This accumulation leads to notable effects on stellar evolution, such as altered thermal transport in main sequence stars~\cite{1985ApJ...294..663S, Iocco:2012wk, Banks:2021sba}. Specifically, ADM interactions, depending on their nature, can modify the temperature gradient in a star's core by evacuating nuclear energy, with~spin-dependent interactions playing a significant role during core hydrogen burning~\cite{Raen:2020qvn}.

% ==========================================
\subsubsection{Influence on Supernovae~Dynamics}
% ==========================================
The deaths of stars are also not free from potential dark sector influences. Specifically, the~study of core-collapse supernovae, which mark the cataclysmic end of massive stars (with masses exceeding approximately $8\, M_{\odot}$) offers a unique window for probing possible hidden sector interactions with ordinary matter. The~emission of light particles from supernova remnants introduces an additional cooling mechanism, which operates by draining energy from the protoneutron star (PNS). This process is subject to constraints, as~it could lead to a reduction in the duration of the neutrino burst observed from SN 1987A, which is a~pivotal discriminant among astrophysical observations~\cite{PhysRevLett.58.1490, PhysRevLett.58.1494, RAFFELT19901}. To~maintain consistency with these observations, a~conservative upper limit on the luminosity of dark particles ($L_d$) has been established, thus ensuring that it remains below the neutrino luminosity observed during the Kelvin--Helmholtz cooling phase, which takes place approximately one second after the core bounce. This constraint results in the formulation of a specific condition: $L_d \lesssim 3 \times 10^{52}\, {\rm erg}\, $s$^{-1}$, which is imposed one second after the bounce~\cite{raffelt1996stars}. Supernova 1987A has also been instrumental in establishing limits on various extensions of the dark sector. These include constraints on axions~\cite{PhysRevLett.60.1793}, dark photons~\cite{Dent:2012mx}, and,~notably, on muonic dark forces~\cite{Croon:2020lrf, Manzari:2023gkt}. The~latter, particularly the gauged $L_{\mu} - L_{\tau}$ model~\cite{PhysRevD.43.R22, Ma:2001md}, has garnered interest for potentially resolving both the muon $(g - 2)_{\mu}$ anomaly~\cite{Aoyama:2020ynm, Muong-2:2021ojo} and~the Hubble constant tension~\cite{Planck:2018vyg, Escudero:2019gzq, DiValentino:2021izs}. Moreover, SN 1987A has been critical in constraining hyperon decay processes, thereby encompassing both decays into light bosons~\cite{Camalich:2020wac} and those leading to GeV-scale dark sector particles~\cite{Alonso-Alvarez:2021oaj} ---this is a constraint that we revisit in our study of neutron dark decay constraints in neutron~stars.

The exploration of supernovae, which is crucial for understanding the dynamics of dark sectors, culminates in the birth of compact star remnants, which leads us to our next topic. 
% ==========================================
\subsection{Neutron Stars as Dark Matter~Laboratories}
\label{subsec:NS_as_DM_Lab}
% ==========================================
Among these remnants, neutron stars stand out due to their extreme densities and strong gravitational fields, thus making them ideal laboratories for probing dark matter effects in dense matter~\cite{Baryakhtar:2022hbu, Bramante:2023djs}. In~this discussion, we categorize the effects of the dark sector on neutron stars into two distinct types: static and~dynamic.

Static effects pertain to modifications in the equation of state (EoS) that governs the dense matter within neutron stars, with~\cite{Berryman:2021wom} as an explicit example. These alterations reshape the neutron stars' configuration space in hydrostatic equilibrium, as~dictated by the EoS. Such effects reach their complete manifestation well within the age (its observation time) of the star, thus acting in a time-independent manner during our~observations.

Dynamic effects, in~contrast, involve changes that occur over time, thereby impacting the neutron star's structure, such as its temperature and spin. These changes evolve with the star, thus reflecting ongoing~transformations. 

It is important to recognize that a new physics model can induce both static and dynamic effects on neutron stars. There are models that exclusively yield dynamic effects while still preserving the original EoS. In~such cases, these models facilitate a quasi-equilibrium transformation of the star, thereby unfolding within a configuration space that is derived from, and~remains consistent with, the~unaltered EoS~\cite{Berryman:2022zic}.

% ----------------------------------------
\subsubsection{Static Effects of Dark Matter on Neutron~Stars}
% ----------------------------------------
Building on the concept of static dark effects, a~notable 
example is the introduction of a dark baryon. This particle, comparable to a baryon in mass but with markedly weaker interactions, results in a decrease in the pressure within neutron stars. Such a phenomenon effectively ``softens'' the EoS. This softening consequently reduces the maximum stable mass for neutron stars ($M_{\rm TOV}$) as determined by the Tolman--Oppenheimer--Volkoff (TOV) equations for hydrostatic equilibrium~\cite{Oppenheimer:1939ne, Tolman:1939jz}. Such modifications present significant challenges to the simplest models aimed at resolving the neutron lifetime anomaly~\cite{Universe:Fornal}, as~we have noted in Section~\ref{sec:intro:NSconstraints}. They reduce $M_{\rm TOV}$ to about $0.7\, M_{\odot}$~\cite{Mckeen2018PhRvL.121f1802M,Baym2018PhRvL.121f1801B,Motta:2018rxp}, which is markedly below the mass of the most massive observed neutron stars, which exceed $2\, M_{\odot}$~\cite{Romani:2022jhd, Fonseca:2021wxt, Antoniadis:2013pzd}. To~reconcile these discrepancies, the~introduction of repulsive interactions either within the dark sector or between dark particles and baryons becomes necessary~\cite{Baym2018PhRvL.121f1801B, Mckeen2018PhRvL.121f1802M, Motta:2018rxp, Grinstein:2018ptl, Universe:NS_Husain, Universe:Zhou}. Alternatively, a~dark decay channel leading to a significantly lower equilibrium density of dark particles could circumvent this stringent bound~\cite{Strumia:2021ybk}.  In~contrast, the~condensation of light scalar fields, such as axions, within~neutron stars—induced by the star's baryonic content—results in a reduction of the baryon masses. This phenomenon can precipitate the emergence of a new ground state in nuclear matter, thereby potentially leading to an increased maximum mass for neutron stars~\cite{Balkin:2020dsr, Balkin:2023xtr}.

% ============================================
\subsubsection{Dark Dynamics of Neutron~Stars}
% ============================================
Dynamical effects on neutron stars can manifest themselves in various forms, ranging from abrupt, dramatic events to more gradual changes. A~prime example of a sudden effect is the collapse of a neutron star into a black hole. Such extreme transformations can be triggered by specific interactions with dark matter, as~seen in three distinct~scenarios.

In one scenario, certain weakly interacting massive particle (WIMP)  variants can accumulate within neutron stars. Over~time, these WIMPs may gravitate towards the star's center, thus eventually forming a self-gravitating core. This class of possibilities includes massive bosons lacking vectorial repulsive interactions and both bosons and fermions with strong attractive scalar interactions. Such a concentration of WIMPs could precipitate the rapid destruction of neutron stars in~a period shorter than their observed lifespans~\cite{PhysRevD.40.3221}. 

\textls[-15]{In a parallel scenario, the~accretion of nonannihilating dark matter within a neutron star can result in the formation of a degenerate dark core. If~this core expands beyond its Chandrasekhar limit~\cite{1931ApJ....74...81C}, the~neutron star hosting it might undergo} 
collapse~\cite{deLavallaz:2010wp, McDermott:2011jp}. This process potentially leads to the creation of anomalously low-mass black holes within~a mass range of approximately $(1$--$2.5)\, M_{\odot}$. Such black holes, if~they merge, could produce gravitational waves detectable by the LIGO-Virgo-KAGRA network. The~absence of detections thus far provides constraints on the interactions of nonannihilating dark matter, as~suggested by recent studies~\cite{PhysRevLett.131.091401}. However, the~possibility of dark matter-induced core collapse can be highly sensitive to dark matter self-interactions~\cite{Bramante:2013hn}. Even extremely weak repulsive self-interactions can prevent black hole formation, which is a~scenario that is particularly relevant in the context of realistic bosonic dark matter models~\cite{Bell:2013xk}. Furthermore, the coannihilation of dark matter with nucleons, which is a~common feature in many asymmetric dark matter models, can also act to inhibit black hole formation~\cite{Bell:2013xk}.  

A third avenue explores the impact of dark matter realized as charged massive particles, with~masses ranging from $10^2\, {\rm GeV}$ to $10^{16}\, {\rm GeV}$. These particles can accumulate in collapsing protostellar clouds. After~the formation of a neutron star, they fall to its center, thereby potentially forming a black hole on a time scale of the order of a year. This black hole could then grow rapidly through accretion, thus leading to the host neutron star's destruction on a similar time scale. The~existence of old neutron stars suggests that these charged massive particles, if~present, make up no more than $10^{-5}$ of the dark halo's mass~\cite{GOULD1990337}.

Gradual dynamical effects can also occur. The~thermal history of neutron stars can deviate from the predictions of standard cooling theory~\cite{Yakovlev:2004yr}. \textls[-15]{This deviation, potentially speaking to new heating mechanisms, has been supported by observations of several old warm neutron stars~\cite{Rangelov:2016syg, Durant_2012}. These stars exhibit surface temperatures higher than those predicted by standard cooling models. They could arise from the inclusion of further SM effects~\cite{Gonzalez_2010,Kopp:2022lru},} including nonequilibrium processes~\cite{Yanagi:2019vrr} such as rotochemical heating, which results from the spin-down behavior %Please check meaning retained 
of pulsars~\cite{Reisenegger:1994be}. However, heating could also be incurred through dark matter scattering and capture~\cite{Baryakhtar:2017dbj,Bell:2018pkk,Bell:2020jou,Kopp:2022lru} or annihilation~\cite{Kouvaris:2007ay}. Alternatively, dark matter effects could alter the structure of dense matter altogether through the appearance of axion quark nuggets~\cite{Zhitnitsky:2023fhs} or of finite density-seeded vacuum instabilities~\cite{Balkin:2021wea,BalkinSciPostPhys}.

There are various heating effects. Kinetic heating, resulting from dark matter capture, involves dark matter particles being gravitationally accelerated to velocities exceeding half the speed of light. As~these particles scatter within neutron stars, they deposit significant kinetic energy, thus possibly heating the stars to infrared black body temperatures~\cite{Baryakhtar:2017dbj}. Although~the dynamics within neutron star cores, the~primary region of such scattering, are not fully understood, the~role of the crust has been emphasized in dark matter interactions. Scattering with just the neutron star's low-density crust can still kinetically heat the star to infrared temperatures, which are detectable by forthcoming telescopes, thereby highlighting the crust's importance in dark matter detection and annihilation processes~\cite{Acevedo:2019agu}. Moreover, inelastic dark matter could significantly influence neutron star heating. Its interactions would be suppressed in standard nuclear recoil direct detection experiments, where small momentum transfers are insufficient relative to the mass splitting between dark matter states. However, neutron stars enable effective inelastic scattering, even with large mass splittings, due to the acceleration of dark matter particles to velocities approaching the speed of light during their infall~\cite{Bell:2018pkk, Alvarez:2023fjj}. Additionally, the~annihilation of WIMPs inside neutron stars can impact their thermal properties. This is particularly evident in neutron stars older than 10 million years that are situated nearer to the Galactic center, where the density of dark matter is expected to be significantly higher than in the vicinity of the Earth. For~instance, in~regions where the dark matter density is at least $3\, {\rm GeV}/{\rm cm}^3$, such interactions could lead to an asymptotic internal temperature of around $10^{4}\,\rm K$~\cite{Kouvaris:2007ay}. Conversely, in~regions with comparatively lower dark matter densities, other heating mechanisms may play a more dominant role in shaping the thermal evolution of neutron stars~\cite{Gonzalez_2010}.

Studies of young neutron stars also give constraints. 
Temperature observations of NS1987A, the~remnant of SN1987A~\cite{Cigan:2019shp, Page:2020gsx}, and~of Cassiopeia A (Cas A), the~second-youngest known neutron star with an approximate age of 340 years, were shown to impose constraints on anomalous cooling processes in neutron stars. These observations play a crucial role in constraining the emissions of novel bosonic particles that could serve as additional cooling mechanisms, thereby altering the thermal evolution of stars~\cite{Hong:2020bxo, Shin:2021bvz}.

As a separate tack, binary pulsar systems offer unparalleled insights into both gravitational physics and particle physics, particularly in the context of dark sectors. These systems enable precision tests of fundamental theories, thereby providing constraints on new physics beyond the SM. In~gravitational physics, binary pulsars have been instrumental in rigorously testing general relativity and alternative theories of gravity in strong-field regimes~\cite{freire2012scalar, stairs2003testing}. Notably, the~Hulse--Taylor pulsar (PSR B1913$+$16)~\cite{hulse1975discovery, Weisberg:2016jye}, in~a binary system of a pulsar and a neutron star, and~the double pulsar system (PSR J0737-3039A/B)~\cite{lyne2004double, PhysRevX.11.041050} have played key roles in validating general relativity's predictions~\cite{PhysRev.136.B1224, PhysRev.131.435}. Furthermore, the~sensitivity of binary pulsar orbital periods to energy loss mechanisms~\cite{1963ApJ...138..471H,10.1093/mnras/85.1.2, 10.1093/mnras/85.9.912} makes them ideal for investigating particle interactions that extend beyond the SM~\cite{Goldman:2009th}. Observations of these periods have led to constraints on ultralight dark matter candidates~\cite{Blas:2016ddr, Armaleo:2019gil, KumarPoddar:2019ceq, Dror:2019uea} and limits on dark baryon processes involving neutron--mirror neutron mixing~\cite{Goldman:2019dbq}, as~well as broader constraints on various forms of the quasi-equilibrium baryon number violation (BNV)~\cite{Berryman:2022zic,Berryman:2023rmh}. The~implications of these constraints, derived from binary pulsar period observations, on~quasi-equilibrium BNV and specific dark decays of baryons are further explored in Section~\ref{sec:BNV:outcome}.

The precision of pulsar timing has become an essential tool in astrophysics, especially through pulsar timing arrays (PTAs), which have been used to find evidence for a stochastic gravitational wave background~\cite{Agazie_2023, EPTA:2023fyk, Reardon_2023, Xu_2023}. This technique also plays a pivotal role in exploring ultralight dark matter (ULDM), as~ULDM candidates with masses around $10^{-22}$~eV have been predicted to create oscillating gravitational potentials at nanohertz frequencies~\cite{Khmelnitsky:2013lxt}. Such oscillations lead to periodic variations in the arrival times of radio pulses from pulsars, thereby overlapping with the frequency range for gravitational waves detectable by PTAs~\cite{porayko2018parkes}. The~European Pulsar Timing Array has effectively used this aspect to  constrain scenarios in which dark matter interacts only gravitationally, thereby demonstrating that ultralight particles in the mass range $10^{-24.0}\, {\rm eV} \lesssim m \lesssim 10^{-23.3}\, {\rm eV}$ cannot fully account for the observed local dark matter density~\cite{PhysRevLett.131.171001}. Additionally, pulsar timing exhibits sensitivity to the effects of heavier dark matter candidates, including primordial black holes and compact subhalos, thereby extending its utility in constraining a wide mass range of dark matter candidates~\cite{Dror:2019twh}. In~addition, pulsar timing shows potential for probing BNV~\cite{Zakeri:2023xyj}. Anomalies in pulsar spin-down rates and braking indices could be indicative of BNV processes, thereby possibly leading to pulsar spin-up behavior and the reactivation of dormant pulsars. With~current timing precision being sensitive to the BNV rates of $\Gamma{\sim}10^{-9}\, {\rm yr}^{-1}$, PTAs are increasingly becoming a crucial tool for future BNV detection or constraints, particularly as their precision improves in the context of gravitational wave~detection.

% ==========================================
\subsection{quasi-equilibrium Dynamics of Baryon Loss in Neutron~Stars}
\label{sec:quasi-eq:NS}
% ==========================================
In Section~\ref{subsec:NS_as_DM_Lab}, we explored the ways in which dark sector extensions to the SM can impact neutron stars. Building on this foundation, this section delves further into these phenomena, with~a particular focus on the implications of specific dark decays of baryons within neutron~stars.

One intriguing example involves neutron dark decays, such as $n\to \chi\gamma$. These should occur at a rate of around $10^{-5}\, {\rm s}^{-1}$ to resolve the neutron lifetime anomaly. If~the dark decay products, such as $\chi$, remain stable over short timescales, they could swiftly saturate the interior of a neutron star shortly after the collapse of its progenitor star's core. This process is expected to drive dynamical effects whose outcomes would be chemical equilibrium between the decay products and the star's baryons. Modeling the EoS under these circumstances necessitates incorporating the new degrees of freedom, such as $\chi$ and other dark sector particles, affecting the star's pressure and energy density. The~ground state of this system would be determined by the lowest energy state, subject to charge conservation and an extended baryon number concept, where $B(\chi) = 1$. Should $\chi$ maintain stability over long timescales, the~neutron star would eventually settle into a static state. This scenario illustrates a static dark sector influence, thereby altering the star's configuration space via its EoS. In~this case, static limits such as the maximum mass of neutron stars can be used to constrain the dark~decays. 

If neutron decay products such as \( \chi \) further decay or engage in annihilation, leading to particles that escape the star, the~situation becomes dynamically complex. It is important to note that these depletion processes need not be direct; they may involve a series of reactions. For~instance, if~$\chi$ decays into leptons, then baryon number conservation would be explicitly violated. This, in~turn, would trigger a sequence of baryon-conserving reactions within the neutron star. The~resulting depletion of baryons manifests itself not only through neutrino emission but also through contributions to the thermal energy and work done on the stellar fluid. In~this case, continuous neutron decays could dynamically influence the neutron star properties over its lifetime. Ongoing processes like these could lead to observable changes in neutron stars, including modifications in thermal evolution, variations in pulsar spins, or~alterations in the orbital periods of binary pulsars. In~this context, neutron stars offer an observational platform for constraining and potentially revealing active dark decay~processes.

\begin{figure}[H]

\includegraphics[width=0.87\textwidth]{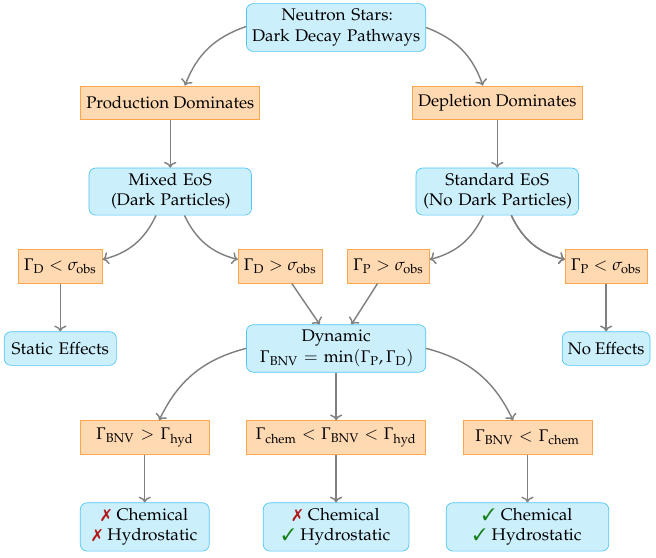}
\caption{ \label{fig:NS_Dark_pathways}
       The potential pathways and resultant states stemming from neutron decays to dark final states in neutron stars, which are dictated by production (\(\Gamma_{\text{P}}\)) and depletion (\(\Gamma_{\text{D}}\)) rates. Observational sensitivity (\(\sigma_{\text{obs}}\)) and the star's response rates—chemical (\(\Gamma_{\text{chem}}\)) and hydrodynamic (\(\Gamma_{\text{hyd}}\)) ones—further influence these pathways. A~departure from chemical and/or hydrostatic quasi-equilibrium (denoted by \textcolor{nicered}{\ding{55}}) necessitates numerical simulations for precise stellar modeling, with~the expectation that stars transition from a nonequilibrium to a quasi-equilibrium state over time. 
        }
\end{figure}

Broadening our perspective, the~overall influence of baryon dark decays, which should dominate dark baryon processes in neutron stars because of the sheer number of neutrons involved,  depends on the production (\( \Gamma_P \)) and depletion (\( \Gamma_D \)) rates of dark particles, as~well as the sensitivity (\( \sigma_{\text{obs}} \)) of neutron star observations. While the depletion mechanisms themselves vary---manifesting as either the direct escape of dark particles from the star or an explicit BNV---the crucial factor in modeling these effects is their rate, with~one notable exception. The~thermal evolution of a neutron star does require a more detailed examination of the depletion mechanism, since it is influenced by the specific sequence of reactions involved in the process. These diverse scenarios necessitate distinct computational approaches, as~depicted in Figure~\ref{fig:NS_Dark_pathways}.

In cases where depletion (\( \Gamma_D \)) exceeds production (\( \Gamma_P \)), the~star primarily comprises SM particles. Here, \( \Gamma_P \) becomes the effective BNV rate to compare against \( \sigma_{\text{obs}} \). Conversely, if~production (\( \Gamma_P \)) surpasses depletion (\( \Gamma_D \)), dark sector particles must be included in the EoS. The~depletion rate (\( \Gamma_D \)) then serves as the effective BNV rate, because~this determines the time scale over which the star~changes.

In scenarios in which the neutron star observations are not sensitive to the BNV rate (\( \Gamma_{\rm BNV} < \sigma_{\text{obs}} \)), we have the following:
\begin{enumerate}[label=\arabic*.]
    \item Production dominance (\( \Gamma_P > \Gamma_D \)) leads to static effects via EoS modifications, thus incorporating dark particles.
    \item Depletion dominance (\( \Gamma_P < \Gamma_D \)) results in no observable effects.
\end{enumerate}

Conversely, when observations are sufficiently sensitive to 
possible BNV effects 
(\( \Gamma_{\rm BNV} > \sigma_{\text{obs}} \)) and~under the assumption that in degenerate cold neutron stars the hydrodynamic response rate exceeds the chemical rate ($\Gamma_{\rm hyd} > \Gamma_{\rm chem}$), we have the following:
\begin{enumerate}[label=\Roman*.,leftmargin=*,labelsep=0.8mm] 
    \item \label{item:quasi-eq} The star is in both chemical and hydrostatic quasi-equilibrium if \( \Gamma_{\rm BNV} < \Gamma_{\rm chem} \). This state warrants a quasi-equilibrium analysis, as~developed in~\cite{Berryman:2022zic}.
    \item The star remains only in hydrostatic quasi-equilibrium if \( \Gamma_{\rm chem} < \Gamma_{\rm BNV} < \Gamma_{\rm hyd} \). A~hydrostatic equilibrium description is valid on short timescales ($\delta t < \Gamma_{\rm BNV}^{-1}$). However, a~complete reaction chain must be considered, as~chemical equilibrium is absent in \linebreak  the EoS.
    \item The absence of both equilibria when \( \Gamma_{\rm BNV} > \Gamma_{\rm hyd} \) requires comprehensive simulations of the star's hydrodynamic and chemical evolution.
\end{enumerate}

% =======================================================
% =======================================================

Notably, in~the quasi-equilibrium scenario (\ref{item:quasi-eq}), baryon number conservation on short timescales forms a key constraint in the derivation of its EOS. The~primary distinction between production- and depletion-dominated states lies in the inclusion or exclusion of dark sector elements and the extended definition of the baryon~number.

The quasi-equilibrium framework posits that if dark sector perturbations, such as baryon decays, proceed at a slower rate than the star's responses, the~star will evolve within an equilibrium configuration space defined by the EoS. In~scenarios where dark particle production dominates, this EoS incorporates the pertinent dark degrees of freedom. Conversely, in~cases where depletion prevails, the~EoS remains akin to the standard formulation, which is exclusive of dark sector influences. This approach effectively balances the dynamical aspects of dark sector interactions against the inherent stabilizing mechanisms of neutron stars. The~computational simplicity in depletion-dominated scenarios is significant, as~the EoS needs not reflect intricate dark sector details, thus facilitating a model-independent~analysis.

Building upon this framework, the~configuration space of neutron stars can be effectively parameterized by a set of \( n \) independent parameters, which is denoted as \( \{\Xi_1, \Xi_2, \ldots, \Xi_n\} \). Variations in the neutron star observables, symbolized as \( \mathcal{O} \), can then be quantified through the following expression:
\begin{equation}
\dot{\mathcal{O}} \equiv \frac{d \mathcal{O}}{dt} = \sum_{i=1}^{n} \left(\frac{d \Xi_i}{dt}\right) \frac{\partial \mathcal{O}}{\partial \Xi_i}, \label{eq:Odot:general}
\end{equation}
where each term \( {\partial \mathcal{O}}/{\partial \Xi_i} \) represents the sensitivity of the observable to changes in the \( i \)th parameter, as~governed by the EoS. The~rates of change for these parameters, \( {d \Xi_i}/{dt} \), are determined by imposing constraints on the system's quasi-equilibrium dynamics. These constraints may include theoretical principles, such as the conservation or specific rate of violation of the baryon number, alongside observational data like the spin-down rates of pulsars. Together, these factors provide a comprehensive framework to model and understand the evolution of neutron star observables within the prescribed parameter~space.

Before exploring specific examples, we think it crucial to recognize that certain neutron star characteristics, such as its surface temperature, rotational period, or~its magnetic field, may not always be central to our framework, particularly when these factors do not significantly influence the observable under study. Neutron stars, which are composed of cold, degenerate matter, experience minimal thermal contributions to their structure. Similarly, rotational effects are often minor for pulsars with longer periods, generally those exceeding a few milliseconds~\cite{Glendenning:1997wn}. Therefore, dynamical evolution within the configuration space can often be treated as independent from changes in the magnetic field, temperature, or~rotation period. However, it is important to note that these elements should be explicitly included in the analysis when they directly impact the observable being studied. For~example, when investigating pulsar spin-down behavior, rotation remains a crucial component within the formalism. Similarly, temperature may be a vital factor in processes that are subject to Pauli blocking~mechanisms. 

We now illustrate the application of the quasi-equilibrium framework with three distinct examples, each of which is succinctly summarized in Table~\ref{tab:ns_quasi-eq_examples}. In~each case, we note what configuration space is modified through the dynamical change indicated, as well as the constraints that act on those~dynamics. 

% =======================================================
\begin{table}[H]
\caption{quasi-equilibrium frameworks for neutron star evolution under various scenarios. Parameters include the central energy densities for the entire star (${\cal E}_c$), visible sector (${\cal E}_c^V$), and~dark sector (${\cal E}_c^D$), as~well as the central angular velocity ($\omega_c$). The~observed spin-down rate is indicated by $\dot{\Omega}$, while the rate of baryon number violation or conversion in dark sector processes is denoted by $\Gamma$.}
\label{tab:ns_quasi-eq_examples}

\begin{tabularx}{\textwidth}{>{\raggedright\arraybackslash}m{3.3cm}>{\raggedright\arraybackslash}m{3.20cm}>{\raggedright\arraybackslash}m{5.95cm}}
\toprule
\textbf{Scenario} & \textbf{Configuration Space} & \textbf{Constraints \& Model Inputs} \\
\midrule
Pulsar Spin-Down & 
$\Xi = \{{\cal E}_c, \omega_c \}$ & 
Observed spin-down rate ($\dot{\Omega}$) \newline Baryon conservation ($\dot{B} = 0$) \\
\midrule
Dark Core Formation & 
$\Xi = \{{\cal E}_c^V, {\cal E}_c^D  \}$ & 
Baryon conversion rate ($\dot{B}_D = \Gamma B_{V}$) \newline Total baryon conservation ($\dot{B}_D = -\dot{B}_V$) \\
\midrule
Baryon Dark Decay 
& 
$\Xi = \{ {\cal E}_c  \}$ & 
Total baryon loss rate ($\dot{B} = -\Gamma B$) \\
\bottomrule
\end{tabularx}
\end{table}
% =======================================================
\vspace{-6pt}

\begin{figure}[H]

\includegraphics[width=0.9\textwidth]{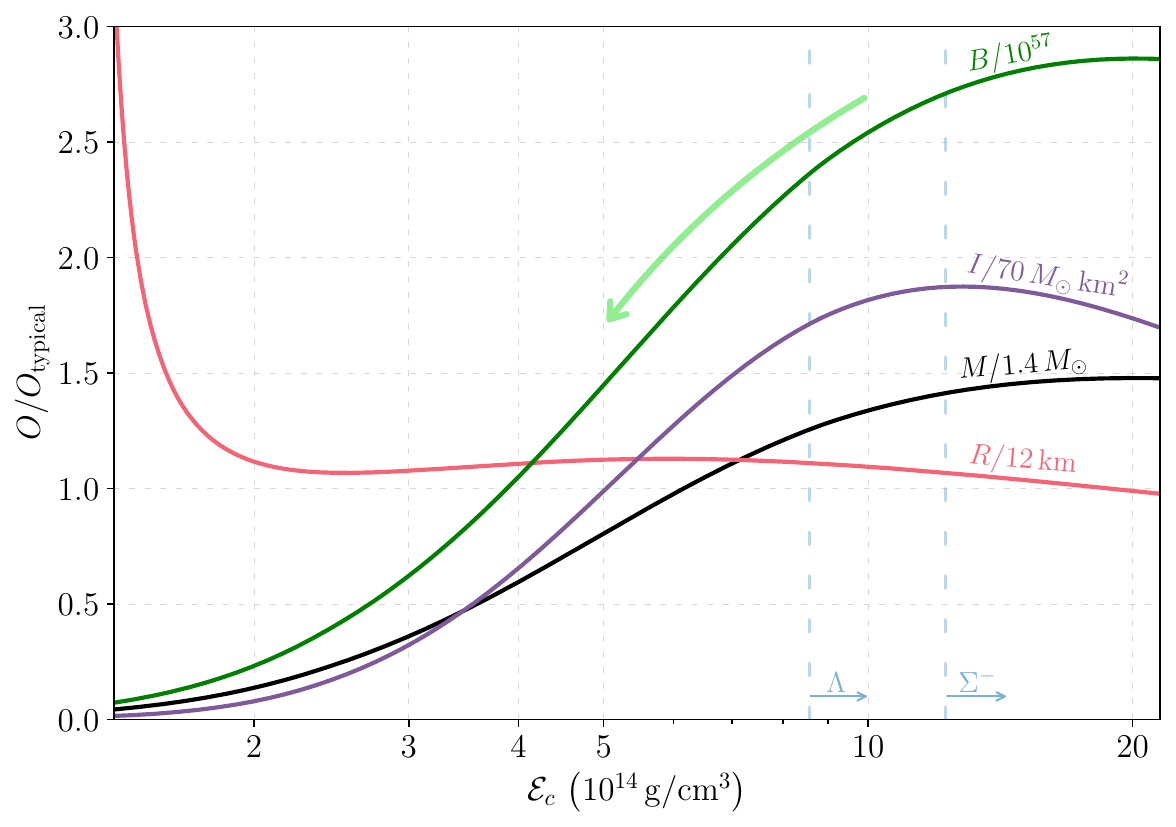}
    \caption{ \label{fig:sequence}
    Variations in neutron star properties as functions of central energy density (${\cal E}_c$): mass ($M$), radius ($R$), baryon number ($B$), and~moment of inertia ($I$), all normalized to typical values. Namely,  $M_{\text{typical}} = 1.4\, M_{\odot}$, $R_{\text{typical}} = 12\, {\rm km}$, $B_{\text{typical}} = 10^{57}$, and~$I_{\text{typical}} = 70\, M_{\odot}\, {\rm km}^2$. The~calculations employ the DS(CMF)-1 EoS~\cite{compose_CMF1}. Notable features include the quasi-equilibrium baryon loss (indicated by the green arrow) and the onset of hyperon appearance (denoted by cyan vertical dashed lines) at sufficiently large central energy density. }
\end{figure}

\begin{enumerate}
    \item Standard Baryon-Conserving Spin-Down Rate of a~Pulsar: %Please check meaning retained
    \begin{itemize}
        \item \textbf{Configuration Space: } This is defined by $\Xi = \{ {\cal E}_c, \omega_c \}$, where 
        ${\cal E}_c$ represents the central energy density, and $\omega_c$ denotes the central angular velocity of the local inertial frames~\cite{Hartle:1967he, Glendenning:1997wn}. 
        \item \textbf{Constraints:} These are governed by the observed spin-down rate ($\dot{\Omega}$) while maintaining baryon number conservation ($\dot{B} = 0$).
    \end{itemize}
    \item Formation of a Dark Core without~Depletion:
        \begin{itemize}
        \item \textbf{Configuration Space:} This is defined by $\Xi = \{{\cal E}_c^V, {\cal E}_c^D \}$, which represents the central energy densities of the visible (${V}$) and dark (${D}$) sectors at the star's core. The~formation rate is generally assumed to be comparable to or slower than the star's age. This rate does not result in thermalization between the two sectors. Consequently, the~sectors are characterized by distinct energy densities and pressures, thereby interacting primarily through gravitational influence.
        \item \textbf{Constraints:} The total baryon number conservation ($\dot{B}_D + \dot{B}_V = 0$) acts as a constraint, thereby ensuring a balanced rate of baryon conversion between the dark and visible sectors. The~conversion rate from visible to dark baryons ($\dot{B}_D$) is derived from the specifics of the particle physics governing the decay or conversion processes.
    \end{itemize}
    \item \label{item:quasi-eq:BNV} 
    Baryon Dark Decays 
    Followed by Efficient~Depletion:
        \begin{itemize}
        \item \textbf{Configuration Space:} This space, also known as the \emph{single-parameter sequence} (illustrated in Figure~\ref{fig:sequence}), is singularly defined by $\Xi = \{ {\cal E}_c \}$ (if the star is not rapidly rotating). Here, ${\cal E}_c$ denotes the central energy density of the neutron star. This simplified configuration space reflects the scenario where baryon dark decays and subsequent depletion processes dominantly influence the neutron star's structure. 
        \item \textbf{Constraints:} The evolution is primarily governed by the total baryon loss rate $\dot{B}$. This rate is instrumental in dictating the evolution of the star's core composition and overall structure, as~indicated by the green arrow in Figure~\ref{fig:sequence}, which represents the quasi-equilibrium evolution along the single-parameter sequence.
    \end{itemize}
\end{enumerate}

% =======================================================
% =======================================================
Transitioning from the comprehensive analysis of neutron stars' quasi-equilibrium dynamics, we now shift our focus to specialized dark sector models, particularly those that expand neutron dark decay scenarios to encompass dark cogenesis. Such models propose a unified origin for the observed baryon asymmetry and a corresponding asymmetry within the dark sector. This approach intriguingly links the physics of neutron stars to wider cosmological~narratives.

\subsection{Illustrative Dark Sector~Models}

We have considered neutron decays in which a dark baryon $\chi$ appears, and~the manner in which we can probe this possibility in a neutron star depends on if and how $\chi$ decays to other dark sector particles. In~this section, we consider just one particular dark sector interaction, but~its impact on a neutron star can vary significantly depending on the assumed spectrum of particles in the dark sector. Indeed, with suitable choices, we will see that it can sample all the scenarios outlined in the previous~section. 

\begin{figure}[H]

    \centering
     \captionsetup{justification=centering}
    \subfloat[\label{fig:darkinduced}]{\includegraphics[width=0.482\linewidth]{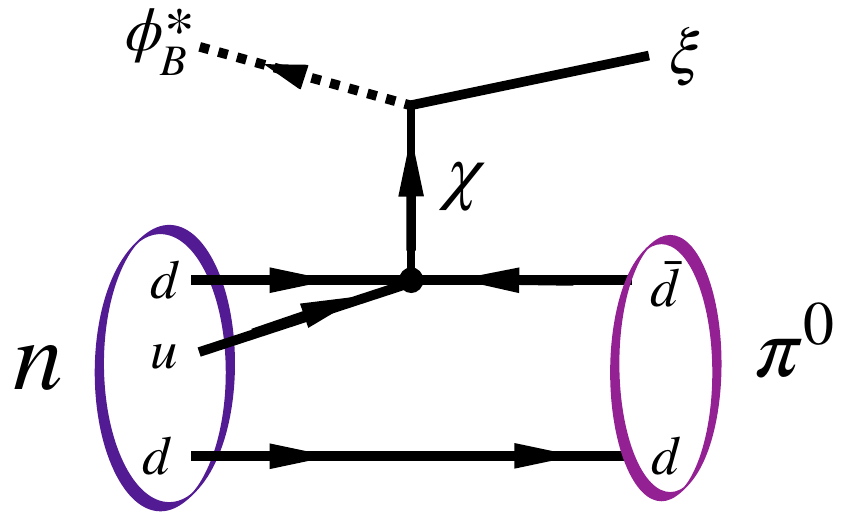}}
    \hfill
     \captionsetup{justification=centering}
    \subfloat[\label{fig:darkann}]{\includegraphics[width=0.50\linewidth]{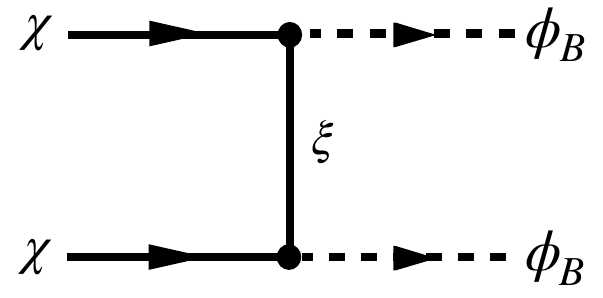}}
    \captionsetup{justification=raggedright}
    \caption{(\textbf{a}) Induced nucleon decay. (\textbf{b}) $\chi\chi$ annihilation from~\cite{Berryman:2023rmh}.\label{fig6}}
\end{figure}

Although we certainly expect $\chi$ to decay were it to play a role in mesogenesis, it need not---perhaps $\chi$ is stable and a possible dark matter candidate in and of itself. This would require, of~course, that the $\chi$ is light enough that it cannot decay to light antibaryons, and~we presume it does not decay just to mesons and/or leptons alone, so that true baryon number-violating ($|\Delta B|=1$) processes do not appear in this context. Thus, as in~\cite{Elor:2018twp,Alonso-Alvarez:2021oaj}, we introduce two new particles: a~scalar baryon $\phi_B$ with $B=1$ and a Majorana fermion $\xi$. It is important that neither $\phi_B$ nor $\xi$ decay, nor that they transform into each other. However, $\chi$ may decay via~\cite{Elor:2018twp}
\begin{equation}
    {\cal L}_{\rm dark} \supset y_d {\bar \chi} \phi_B \xi + \rm h.c. \,,
    \label{eq:chidark}
\end{equation}
where either $\phi_B$ or $\xi$ or both could be dark matter candidates. This interaction can destabilize the proton, produce $\chi$ decay, or produce $\chi\chi$ annihilation, with~examples as shown in Figure~\ref{fig6}a,b. For~different choices of the particles' masses, different processes appear. For~example, we have the following: 

\begin{itemize}
    \item If $M_{\phi_B},M_\xi > M_\chi$, then $\chi$ does not decay. 
    \item If $M_{\phi_B} + M_n > M_\xi + M_{\pi^0}$, then induced nucleon 
    decay ($\phi_B^* + n \to \xi + \pi^0$) 
    can appear, as in Figure~\ref{fig6}a. Similarly, if 
    $M_\xi + M_n > M_{\phi_B} + M_{\pi^0}$, then $\xi + n \to \phi_B + \pi^0$
    can appear. 
    \item If $M_\chi > M_{\phi_B}$ but $M_\xi > M_\chi$, 
    then $\chi\chi \to \phi_B \phi_B$ can occur, 
    as shown in Figure~\ref{fig6}b, though~
    $\chi\to \phi_B \xi$ cannot. Similarly,  if~
    $M_{\chi} > M_{\xi}$ but $M_{\phi_B} > M_\chi$, then $\chi\chi\to \xi\xi$
    can occur instead. 
    \item If $M_\chi > M_{\phi_B} + M_{\chi}$, then 
    $\chi\to \phi_B \xi$ can occur. 
\end{itemize}
If $\chi$ does not decay, the~neutron dark decay rate is constrained, or~alternatively, $\chi$ self-interactions must be present, as~previously noted in Section~\ref{sec:introduction}, to~align with observed neutron star masses. This scenario corresponds to the production dominance route depicted in Figure~\ref{fig:NS_Dark_pathways}, thereby leading to static effects such as the modification of the maximum theoretical neutron star~mass.

However, with~a suitable choice of dark sector masses, subsequent to neutron decay to a final state with a $\chi$ particle, either $\phi_B$ and/or $\xi$ can appear. These outcomes can act with different effects within a neutron star. Either sort of particle could fall to the core of the star and~modify its evolution---and possibly act with a destabilizing effect~\cite{McDermott:2011jp}. Alternatively, if~$\phi_B$ and $\xi$ are light enough, they can escape the star. Thus, the scenarios described in the previous section can all be sampled through the dark sector interaction of Equation~(\ref{eq:chidark}). In~our own investigations~\cite{Berryman:2022zic,Berryman:2023rmh}, we have used a quasi-static approach to determine the evolution of the star in the presence of a slow apparent BNV to dark final states. The~consistency of this analysis, that new physics dynamics does not impact the evolution of the star, requires that we have an effective mechanism to remove energy from the star upon decay to a dark final state. Accordingly, our analysis is predicated on the assumptions inherent in the depletion dominance pathway of Figure~\ref{fig:NS_Dark_pathways}, which culminates in a quasi-equilibrium evolution of the star. As~a concrete example of such, we have supposed that $M_\xi \gg M_{\phi_B} , M_{\chi}$, thus realizing $\chi$ removal through $\chi\chi \to \phi_B \phi_B$ decay, with~$\phi_B$ being light enough to escape the~star. 

Additional physical constraints appear if we work within the context of a dark cogenesis model. For~example, if~$\xi \to \phi_B \chi$ could occur, and~it would under the conditions we have just outlined for quasi-static evolution of the star, then if $\chi$ is heavy enough, $\chi \to {p} \pi^-$ and $\chi \to {n} \pi^0$ can also appear. In~a broader context, such processes would act to dilute a cosmic baryon asymmetry. Thus, either the flavor structure of the pertinent quark--$\chi$ couplings must be sufficiently small, or~the mass of $\chi$ is further constrained. Following the latter path, we note that successful baryogenesis within the dark sector scenario of Equation~(\ref{eq:chidark}) (and masses such that $\chi\chi \to \phi_B\phi_B$ occurs) can be realized if  $m_\chi \le 1.07784 \,\rm GeV$. This and nuclear stability considerations gives us the following window on $\chi$~\cite{Berryman:2023rmh}:
\begin{equation}
0.937993 \,{\rm GeV} < m_{\chi} < 1.07784 \,{\rm GeV} \,,
\label{eq:chimasswindow}
\end{equation}
though in what follows, we consider neutron star constraints on $\chi$--baryon mixing over a larger mass window in~anticipation of other model building solutions. We now turn to the evaluation of these decays within the dense medium of a neutron~star. 

% =======================================================
% =======================================================

% ========================================================
\section{Neutron Stars and Particle Decays in Dense~Matter}
\label{sec:medium}
% ========================================================
The consideration of dense matter phenomena is crucial for investigating baryon interactions in neutron stars, where intense gravitational forces compress baryonic matter to densities exceeding nuclear saturation. Such extreme conditions push the system into the strongly repulsive domain of nuclear interactions, thus resulting in an elevation of the ground state energies of the baryons. This increase in energy levels potentially allows for decay processes that are otherwise kinematically forbidden in both a vacuum and within nuclei~\cite{Berryman:2023rmh}. Consequently, neutron stars present themselves as unique astrophysical laboratories, thereby offering the possibility to observe baryon dark decays that bypass the constraints usually encountered in earthbound experiments such as Super-Kamiokande or  KamLAND~\cite{Super-Kamiokande:2015pys, KamLAND:2005pen}. To~gain a more comprehensive and quantitative insight into this phenomenon, we delve into a particular formalism of dense matter. This exploration aims to uncover the unique role of heavy neutron stars as cosmic laboratories, thus offering new probes of dark~sectors.

In this vein, a~hadronic relativistic mean field (RMF) theory offers a robust framework for addressing dense matter interactions in such extreme environments~\cite{Walecka:1974qa,Serot:1984ey,Serot:1997xg,Dexheimer:2008ax}. Within~RMF theory, baryon interactions are mediated through various Lorentz scalar and vector mesons. The~Euler--Lagrange equations that govern the dynamics of these meson fields include baryon currents as source terms, thus taking the form of $\overline{\psi} \psi$ for scalar mesons and $\overline{\psi} \gamma^{\mu}\psi$ for vector mesons. As~the baryon density increases, these currents intensify, thereby leading to a significant vacuum expectation value (VEV) for the meson fields. Within~this framework, the~mesonic degrees of freedom are thus effectively represented by their classical VEVs, which is a~simplification that sees scalar meson VEVs contributing to the baryon masses and vector mesons to their four-momentums. This theoretical construct provides the foundation for understanding baryon behavior in the dense, locally uniform medium of neutron~stars.

\begin{figure}[H]

\includegraphics[width=\textwidth]{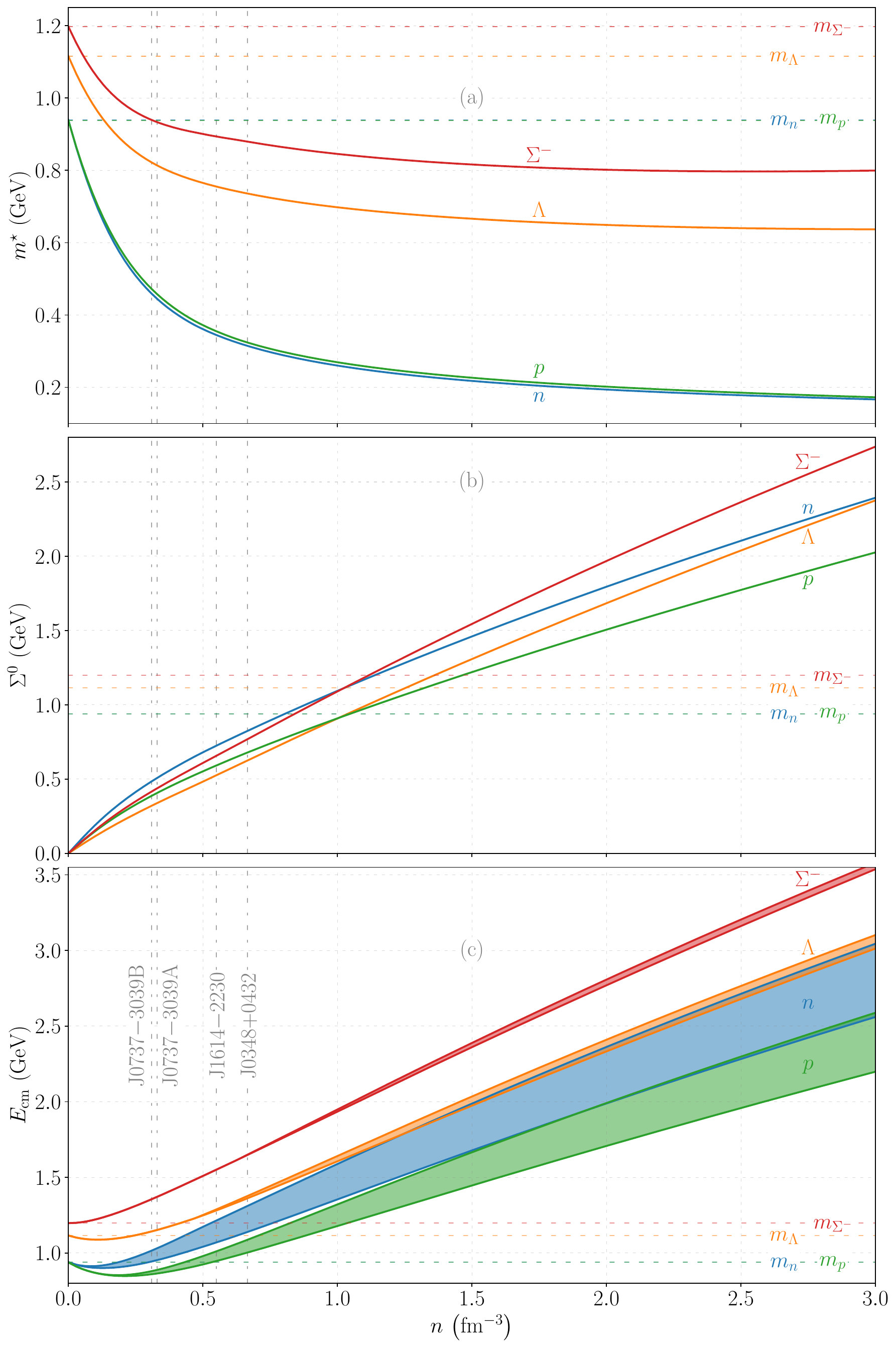}
 \caption{ \label{fig:Micro}
Variations in baryonic properties under the DS(CMF)-1 EoS depicted as functions of density. (\textbf{a}) Effective masses of various baryons. (\textbf{b}) Vector self-energies of the baryons in the nuclear matter frame. (\textbf{c}) Range of center of mass (CM) frame energies for the baryons. Vertical lines in all panels represent the central number densities ($n_c$) of the pulsar systems examined in Section~\ref{sec:BNV:outcome}, while horizontal lines correspond to the vacuum masses of the respective baryons.
}
\end{figure}

In such a medium, the~wave function for a baryon with a canonical four-momentum $p^{\mu}$ is expressed as
\begin{equation}
    \psi (x) = e^{-ip\cdot x} u(p^*, \lambda),
    \label{eq:medium:MFT:modspinor}
\end{equation}
where $p^{*\mu} \equiv p^{\mu} - \Sigma^{\mu} = \left\{E^*(p^*), \vec{p} - \vec{\Sigma}\right\}$ represents the {\it kinetic} four-momentum. The~vector self-energy ($\Sigma^{\mu}$), generated by the vector meson VEVs, is characterized by $\vec{\Sigma} = 0$ in the nuclear matter frame. The~time component of $p^{*\mu}$ is given by $E^*(p^*) \equiv \sqrt{{m^*}^2 + |\vec{p}^*|^2}$, with~$m^*$ signifying the baryon's effective mass. The~baryon spinor $u(p^*, \lambda)$ satisfying the Dirac equation
\begin{equation}
    \left(\slashed{p}^* - m^*\right) u(p^*, \lambda) = 0, \label{eq:medium:MFT:dirac}
\end{equation}
is solved in the Dirac--Pauli representation as
\begin{equation}
    u(p^*, \lambda) = \sqrt{E^*(p^*) + m^*}  
    \begin{pmatrix}
    1 \\
    \frac{\vec{\sigma}\cdot \vec{p}^{\,*}}{E^*(p^*) + m^*}
    \end{pmatrix} 
\eta_{\lambda},\label{eq:medium:MFT:spinor}
\end{equation}
where $\vec{\sigma}$ are the Pauli matrices, and~$\eta_{\lambda}$ is the Pauli spinor with $\eta_{\uparrow} = (1, 0)^T$ and $\eta_{\downarrow} = (0, 1)^T$. The~baryons' energy spectrum ($p^0$) is delineated by
\begin{equation}
    E(p) = \sqrt{{m^*}^2 + |\vec{p} - \vec{\Sigma}|^2} + \Sigma^0, \label{eq:medium:MFT:dirac:energy}
\end{equation}
where we note that in the RMF approximation, $\Sigma^{\mu}$ and $m^*$ are density-dependent but independent of $p^{\mu}$. The~CM energy of the baryons, denoted as \( E_{\rm cm}(p) \), is a function of their effective mass \( m^* \) and momentum \( \vec{p} \) in the nuclear matter frame. It is expressed by the following equation:
\vspace{-10pt}
\begin{align}
    E_{\rm cm}(p) = \sqrt{\left[E(p)\right]^2 - \left|\vec{p}\right|^2 } = \sqrt{\left(\Sigma^0\right)^2 + \left(m^*\right)^2 + 2 \Sigma^0 \sqrt{\left(m^*\right)^2 + \left|\vec{p}\right|^2 }}. \label{eq:ECM}
\end{align}
This energy reaches its minimum when the momentum \( |\vec{p}| \) is zero and reaches its maximum at the Fermi momentum. The~numerical values of $m^*$, $\Sigma^{0}$, and~$E_{\rm cm}$ for the baryons within neutron stars under the DS(CMF)-1 EoS~\cite{compose_CMF1}, which comes from a more sophisticated descendent~\cite{Dexheimer:2008ax} of the RMF model we have noted, are depicted as functions of density in Figure~\ref{fig:Micro}. (We anticipate further evolutionary development of the RMF approach through a determination of its parameters from a description of pure neutron matter within chiral effective theory~\cite{Alford_2022PhRvC.106e5804A}.) As evidenced in Figure~\ref{fig:Micro}, a~notable reduction in the effective baryon masses ($m^*$) at higher densities is observed, which is indicative of chiral symmetry restoration in these high-density regimes in this model. Furthermore, the~dynamics between $m^*$ and $\Sigma^{0}$ (in the nuclear matter frame) are such that while $m^*$ decreases, $\Sigma^{0}$ increases, thereby cumulatively leading to an increase in the total energy of the baryons as delineated by Equation~\eqref{eq:medium:MFT:dirac:energy} at higher densities. The~core (maximum) density of four pulsars (in binary systems) are denoted by vertical dot--dashed lines for comparison. This sets the stage for understanding how each neutron star, with~its unique density profile, represents a distinct astrophysical~laboratory.

\begin{figure}[H]

\includegraphics[width=\textwidth]{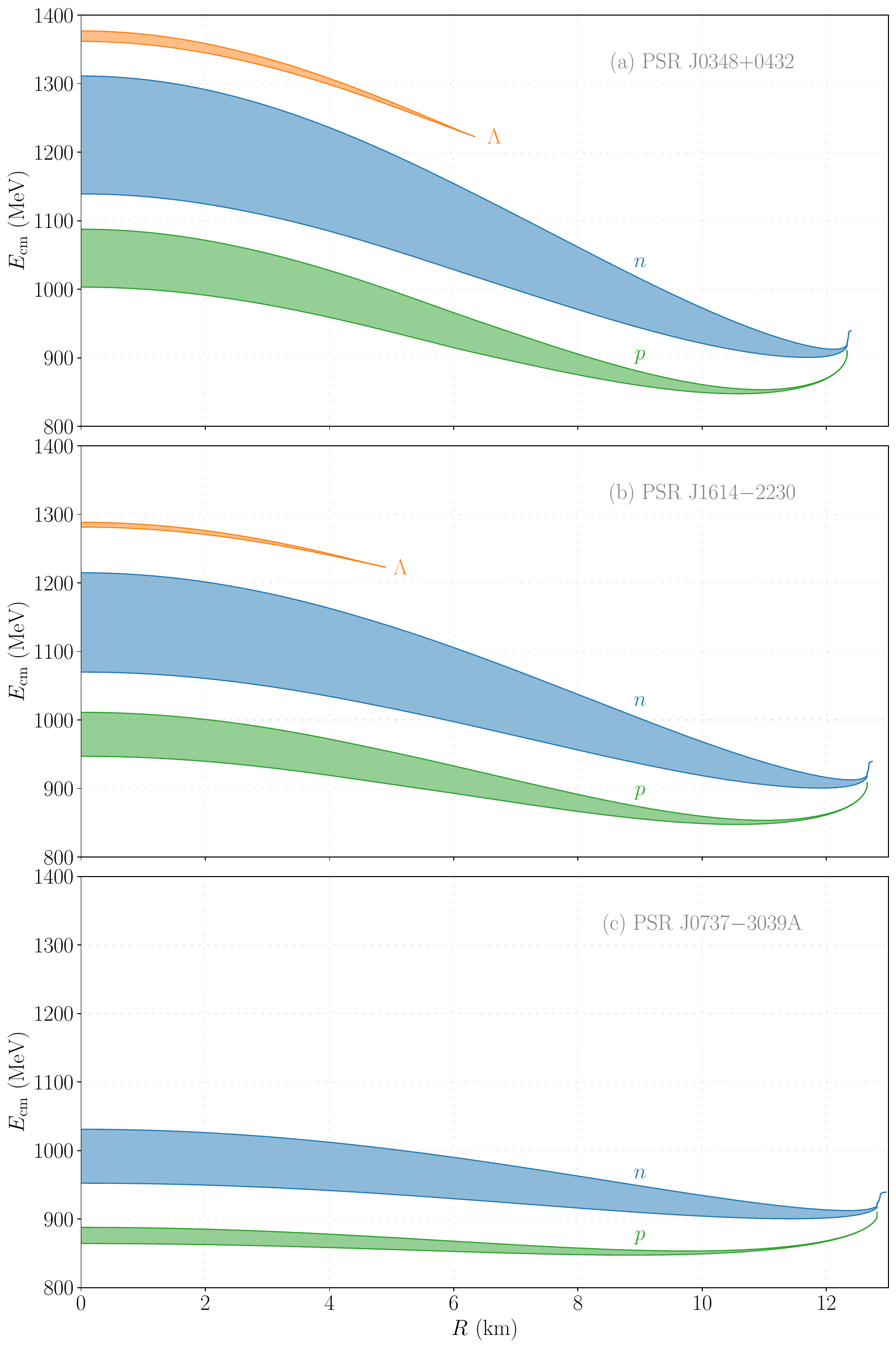}
 \caption{  \label{fig:ECM}
 Energy spectra of baryons ($\Lambda$, neutron ($n$), and~proton ($p$)) in their center of mass (CM) frames within specific neutron stars of (\textbf{a}) PSR J0348+0432, (\textbf{b}) PSR J1614-2230, and (\textbf{c}) PSR J0737-3039A plotted as functions of stellar radius. We have chosen to represent only PSR J0737-3039A here and henceforth, as~it is nearly identical in mass to PSR J0737-3039B. The~plot highlights how the CM energy in high-density regions exceeds the baryons' vacuum rest mass, thus hinting at the possibility of decay into states that are kinematically forbidden in vacuum. Detailed analysis of these phenomena is provided in the main text.
    }
     
\end{figure}

Figure~\ref{fig:ECM} further demonstrates how each neutron star offers a unique probe that is intrinsically linked to its specific density profile. The~CM energy spectrum of the baryons ($\Lambda$, neutron ($n$), and~proton ($p$)) in neutron stars in PSR J1614-2230, PSR J0348+0432, and~PSR J0737-3039A has been observed to surpass their vacuum rest mass within the stars' cores, as~shown in the figure. Notably, heavier neutron stars such as in PSR J1614-2230 and PSR J0348+0432 have been shown to exhibit a higher energy density reach, thereby underscoring the relationship between a star's mass and its potential for exploring baryon dark decays. This characteristic of neutron stars, particularly the more massive ones, highlights their capability to facilitate baryonic decays into states that are kinematically forbidden in vacuum conditions. The~unique energy reaches offered by these pulsar systems offer invaluable insights into dark matter models beyond the scope of terrestrial experiments. In~the ensuing section, we demonstrate how observations of binary pulsars, in~conjunction with the considerations of dense matter outlined herein, pave the way for novel constraints on the dark decays of~baryons.

% =======================================================
% =======================================================

% =======================================================
% =======================================================

% ========================================================
\section{Baryon Loss Constraints from Energy Loss Limits in Pulsar~Binaries}
\label{sec:BNV:outcome}
% ========================================================
Binary pulsar orbital periods, which are sensitive to energy loss mechanisms, present an effective observational tool for probing quasi-equilibrium BNV processes. In~scenarios with efficient depletion, such as Example~\ref{item:quasi-eq:BNV} in Section~\ref{sec:quasi-eq:NS}, the quasi-equilibrium BNV at a rate $\dot{B}$ contributes to an energy loss $\dot{M}^{\rm eff}$ from neutron stars as follows~\cite{Berryman:2022zic}:
\begin{equation}
\label{eq:macro_bnv:observables:mdot:eff}
    \dot{M}^{\rm eff} = \left( \partial_{{\cal E}_c} M +  \left(\frac{\Omega^2}{2}\right) \partial_{{\cal E}_c} I \right) \left(\frac{\dot{B}}{\partial_{{\cal E}_c} B} \right),
\end{equation}
where $\partial_{{\cal E}_c}$ represents the partial derivative with respect to the central energy density of stars, $M$ is the mass, $I$ denotes the moment of inertia, and~$\Omega$ is the angular spin frequency. This energy loss affects the binary pulsar orbital period~\cite{1963ApJ...138..471H,10.1093/mnras/85.1.2, 10.1093/mnras/85.9.912}:
\begin{equation}
 \label{eq:macro_bnv:observables:jeans:mloss}
    \dot{P}_b^{\dot{E}} = - 2 \left(\frac{\dot{M}_1^{\rm eff} + \dot{M}_2^{\rm eff}}{M_1 + M_2}\right)\, P_b,
 \end{equation}
 in which $1$ and $2$ represent binary components. The~observed orbital period decay rate, \(\dot{P}_b^{\rm obs}\), combines gravitational radiation (\(\dot{P}_b^{\rm GR}\)), intrinsic energy loss (\(\dot{P}_b^{\dot{E}}\)), and~extrinsic factors (\(\dot{P}_b^{\rm ext}\)):
\begin{equation}
\dot{P}_b^{\rm obs} = \dot{P}_b^{\rm GR} + \dot{P}_b^{\dot{E}} + \dot{P}_b^{\rm ext}.\label{eq:binary_period}
\end{equation}
The precision in these terms determines the sensitivity of the BNV limits, thus making binary pulsars with accurate timing ideal for realizing stringent constraints. Moreover, as~noted in Section~\ref{sec:medium}, each pulsar, especially the more massive ones, offers a unique opportunity for the study of the BNV. Their larger mass extends the density reach, thereby enabling constraints on higher mass ranges of dark particles such as  $\chi$ in processes such as $n \to \chi \gamma$. Furthermore, selecting binary systems with heavier pulsars is essential for constraining the dark decays of hyperons. In~EoSs that incorporate hyperon degrees of freedom, hyperons start populating the medium only in extremely dense environments~\cite{1960SvA.....4..187A}, specifically in the cores of heavier neutron stars (as illustrated in Figure~\ref{fig:ECM}). 

In our pursuit of comprehensive coverage of the model parameter space, a~strategic combination of binary pulsars---both heavy and precisely measured---is essential. Ideal candidates are systems with accurately determined masses, orbital periods, and~well-quantified contributions to each term in Equation~\eqref{eq:binary_period}. It is critical to select binary systems that are free from mass transfer between their components to avoid complex alterations in the observed orbital period decay rate, as~mass transfer can significantly influence \(\dot{P}_b^{\text{obs}}\)~\cite{1986Natur.319..383R}. Excluding the heavy candidate PSR J0952-0607, a~``black widow'' pulsar with substantial mass uncertainty \(M_p = (2.35 \pm 0.17)\, M_{\odot}\) and significant rotational effects on its structure~\cite{Romani:2022jhd}, our attention turns to the following candidate binary pulsar~systems: 
\begin{itemize}
    \item PSR J0740+6620: \(M_p = (2.08 \pm 0.07)\, M_{\odot}\)~\cite{Fonseca:2021wxt},
    \item PSR J0348$+$0432: \(M_p = (2.01 \pm 0.04)\, M_{\odot}\)~\cite{Antoniadis:2013pzd},
    \item PSR J1614-2230: \(M_p = (1.908 \pm 0.016)\, M_{\odot}\)~\cite{Arzoumanian:2017puf}.
\end{itemize}
We ultimately selected PSR J0348$+$0432 and PSR J1614-2230, which offer the advantages of smaller errors in \(M_p\) and \(\dot{P}_b^{\text{obs}}\), thus ensuring more precise constraints. 
Complementing this selection, we also considered pulsar systems that are renowned for their high-precision measurements. Among~these, the~double pulsar J0737-3039A/B~\cite{PhysRevX.11.041050} and the Hulse--Taylor pulsar system~\cite{hulse1975discovery} stand out. Since both are characterized by similar mass profiles, we focused on the double pulsar J0737-3039A/B, which was chosen for its superior precision. 

\begin{figure}[H]

\includegraphics[width=\textwidth]{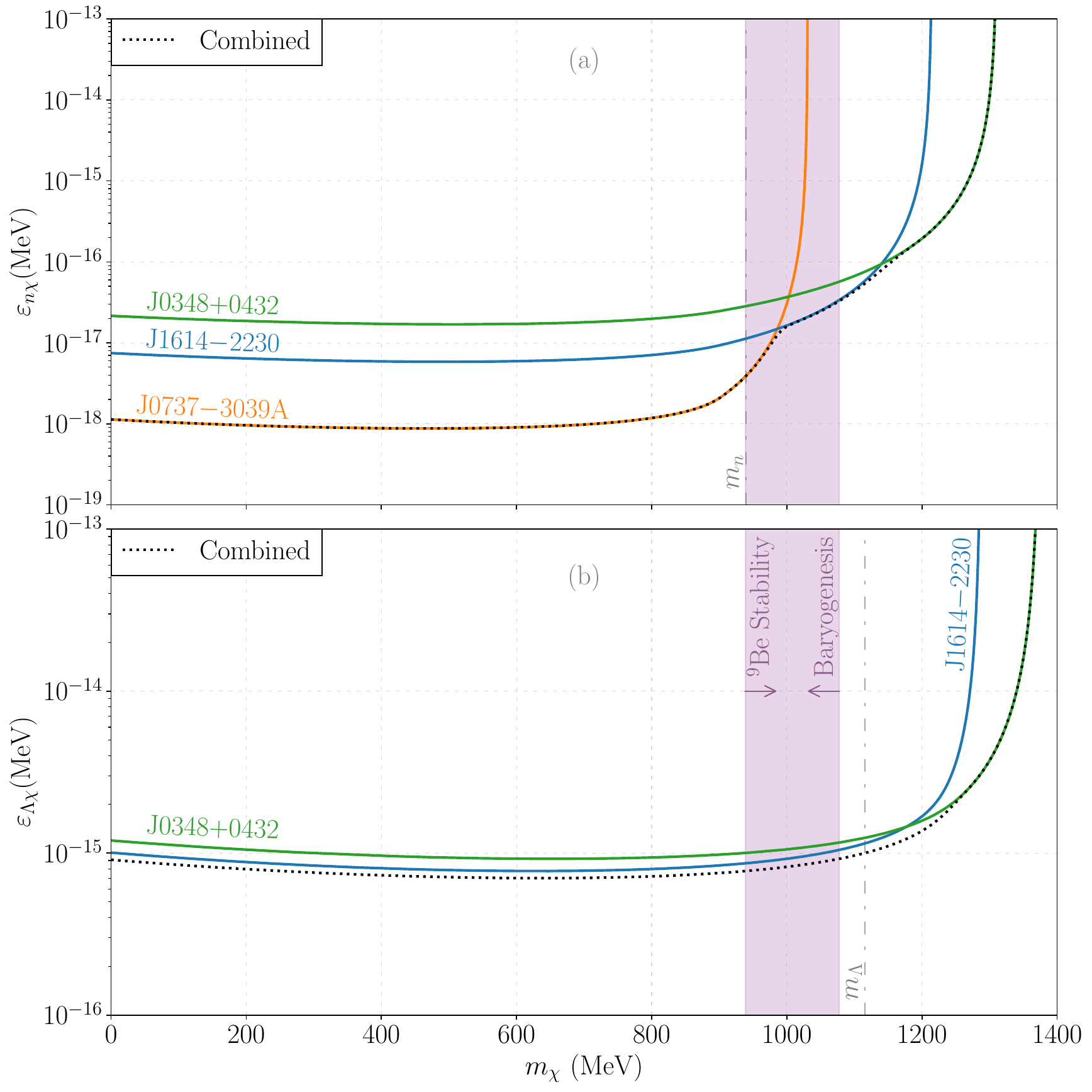}
 \caption{\label{fig:Limits_Combined}
    Two-$\sigma$ exclusion constraints on the mixing parameters for baryon--dark baryon interactions derived under the depletion dominance scenario~\cite{Berryman:2023rmh} as illustrated in Figure~\ref{fig:NS_Dark_pathways} and based on the DS(CMF)-1 EoS~\cite{Berryman:2023rmh}.
    The vertical purple band indicates the range for the dark baryon mass (\(m_{\chi}\)), which is essential for nuclear stability and successful baryogenesis as defined in Equation~\eqref{eq:chimasswindow}. (\textbf{a}) Neutron (\(n\))--dark baryon (\(\chi\)) interaction constraints as functions of \(m_{\chi}\). The~colored lines represent the exclusion limits from the pulsars in PSR J0348+0432 (green), PSR J1614-2230 (blue), and~pulsar A in the double pulsar system PSR J0737-3039A/B (orange), with~the combined limits depicted by a dotted black curve. The~vertical dashed lines mark the vacuum rest mass of the neutron. (\textbf{b}) Similar constraints for the \(\Lambda\)--\(\chi\) interaction using the same color scheme and presentation. Here, the~vertical dashed lines represent the vacuum rest mass of the \(\Lambda\) hyperon.
}
\end{figure}

The constraints derived from our selected pulsar systems were applied to a specific model involving baryon decays \({\cal B} \to \chi \gamma\), which was dictated by an effective mixing parameter \(\varepsilon_{{\cal B} \chi}\). The~in-medium Lagrangian governing these decays is represented as:

\vspace{-6pt}
\begin{equation}
     \mathcal{L} = \overline{\psi}_{\cal B} \left( i 
     \slashed{\partial}
     - \slashed{\Sigma}_{\cal B} - m_{\cal B}^* \right) \psi_{\cal B} + \overline{\psi}_\chi \left(  i 
     \slashed{\partial}
     - m_{\chi} \right) \psi_{\chi} - \varepsilon_{{\cal B} \chi} \left( \overline{\psi}_{\cal B} \psi_{\chi} + \overline{\psi}_{\chi} \psi_{\cal B}\right) \,. \label{eq:dark_decay:medium:method:toy_mix_L}
 \end{equation}
In this equation, we have employed $\slashed{\partial} \equiv \gamma_{\mu} \partial^{\mu}$, where $\gamma_{\mu}$ are the Dirac gamma matrices, and $\partial^{\mu}$ represents the spacetime derivatives. Furthermore, $\psi_{\cal B}$ and $\psi_{\chi}$ correspond to the field operators for visible and dark baryons, respectively. To~construct a combined limit (\(\varepsilon^{\rm comb}_{{\cal B} \chi}\)) from the individual pulsar constraints ($\varepsilon^{\, i}_{{\cal B} \chi}$), we utilize the following relation:
\begin{equation}
    \varepsilon^{\rm\, comb}_{{\cal B} \chi} = \Big[ \sum_i \big(\varepsilon^{\, i}_{{\cal B} \chi}\big)^{-4} \Big]^{-1/4}.
\end{equation}
The resulting limits, as~derived in~\cite{Berryman:2023rmh}, are depicted for individual systems and their combined limits under the assumption of the DS(CMF)-1 EoS~\cite{compose_CMF1} from~the hadronic EoS of~\cite{Dexheimer:2008ax} with a crust~\cite{Gulminelli:2015csa} from the CompOSE database~\cite{CompOSECoreTeam:2022ddl} in Figure~\ref{fig:Limits_Combined}. 
%
% =======================================================
% =======================================================

As depicted in Figure~\ref{fig:Limits_Combined}, the pulsars provided constraints on models with dark baryon $\chi$ masses extending several hundred MeV above the mass of the decaying baryon. Specifically, the~heaviest pulsar, in~PSR J0348$+$0432, demonstrated the broadest energy reach, while the most precisely measured system, the~double pulsar J0737-3039A/B, yielded the most stringent limits for lower $\chi$ mass ranges.
By varying the choice of EoS, we can explore the robustness of our assessments. We have explored this for the entire family of DS(CMF) EoSs: DS(CMF)-1-8~\cite{compose_CMF1,compose_CMF8} in~\cite{Berryman:2023rmh}. Interestingly, the~combined $\varepsilon_{n\chi}$ constraint changed negligibly below ${\sim}1\,\rm  GeV$, thus effectively covering our window of interest, Equation~(\ref{eq:chimasswindow}), \textls[-15]{whereas variations in $\varepsilon_{\Lambda \chi}$ of a factor of a few appeared over the entire mass range studied---and only} DS(CMF)-1,3,7 supported the appearance of strange baryons for the parameters of our candidate pulsars. We note that the $\epsilon_{n\chi}$ parameter at $2\sigma$ was bounded to be no larger than $2\times 10^{-17}$ for the entire window in the $\chi$ mass we have supposed in a dark cogenesis~scenario.

We consider the implications of these results in the next~section. 

% ========================================================
\section{Perspectives on the Neutron Lifetime Anomaly and Dark~Cogenesis}
\label{sec:perspective}
% ========================================================

\begin{figure}[H]

\includegraphics[width=\textwidth]{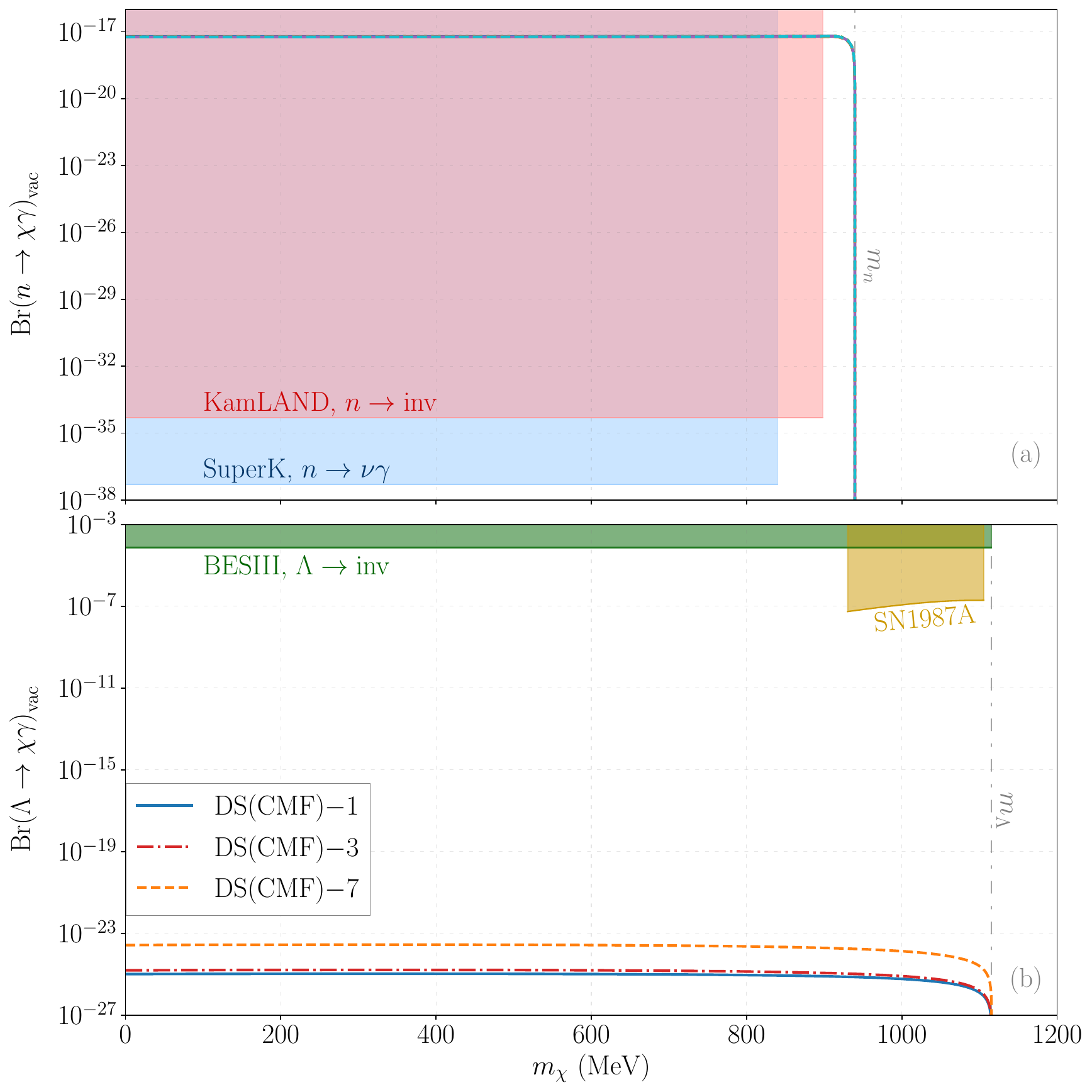}
    \caption{ \label{fig:BR_Limits} Two-$\sigma$ exclusion limits on the vacuum branching fraction for ${\cal B}\to\chi\gamma$ processes are depicted in panels (\textbf{a}) for neutrons and (\textbf{b}) for $\Lambda$ baryons. The~green region represents limits from Super-Kamiokande~\cite{Super-Kamiokande:2015pys}. We highlight that the KamLAND constraint (red)~\cite{KamLAND:2005pen} is pertinent to our study of $n\to \chi\gamma$ decay, as~the prompt photon would not pass the correlation cuts and would remain undetected. Panel (\textbf{b}) also presents limits from neutron star studies for the DS(CMF)-7, DS(CMF)-3, and~DS(CMF)-1 EoSs in order of increasing severity, along with constraints from the BES-III experiment (blue)~\cite{BESIII:2021slv}, noting that the $\gamma$ in this case would also be undetected, and~the SN1987A limits (gold)~\cite{Alonso-Alvarez:2021oaj}.}
\end{figure}

We now determine the outcomes from our neutron star analysis in regard to the in-vacuum limits on neutron dark decay. We can translate the constraints on \(\varepsilon_{{\cal B} \chi}\) that we report in Figure~\ref{fig:Limits_Combined} to ones on the in-vacuum decay rate of \({\cal B}\to \chi\gamma\) using the following equation:
\begin{align}
\label{eq:dark_decay:vacuum:vacdec}
    \Gamma \left({\cal B}\to \chi\gamma\right)
    = \frac{g_{{\cal B}}^2\, e^2\, \varepsilon_{{\cal B} \chi}^2}{128\pi} \frac{\left(m_{{\cal B}} + m_{\chi}\right)^2}{m_{{\cal B}}^5} \left( m_{{\cal B}}^2 - m_{\chi}^2\right). 
\end{align}
We note that \(g_{\cal B}\) signifies the baryon magnetic moment (Land{\'e}) $g$ factor, where \(g_n \simeq 3.826\) for neutrons, and \(g_\Lambda \simeq -1.226\) for $\Lambda$ baryons~\cite{ParticleDataGroup:2022pth}. The~resulting constraints on the vacuum branching ratios are illustrated in Figure~\ref{fig:BR_Limits}.

\begin{figure}[H]

\includegraphics[width=\textwidth]{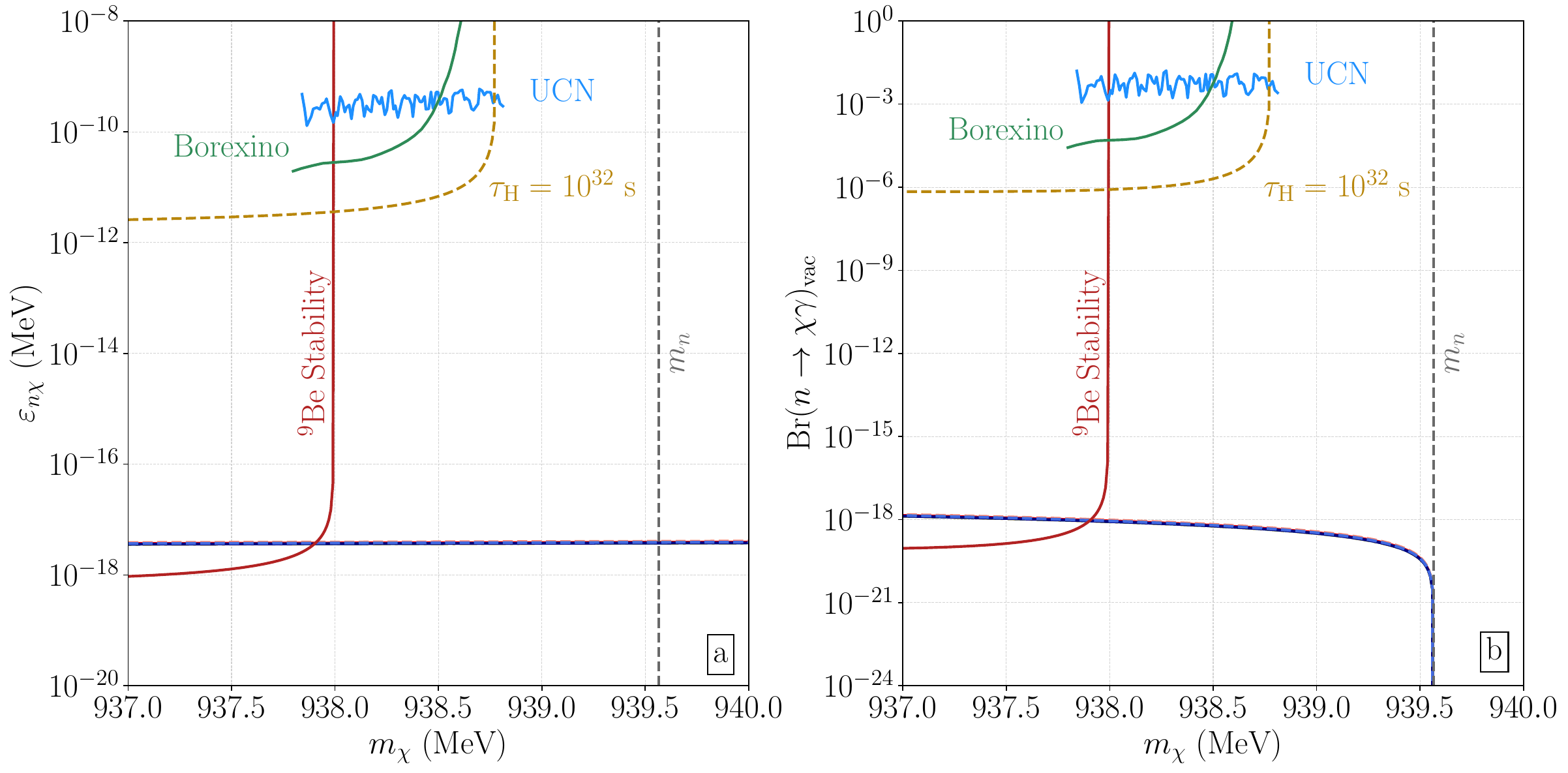}
    \caption{\label{fig:narrow}
    Exclusion limits at 2 $\sigma$ on (\textbf{a}) $\varepsilon_{n\chi}$ and (\textbf{b}) the vacuum branching fraction for $n \to \chi \gamma$ from~our quasi-static neutron star analysis in the particular $\chi$ mass region pertinent to an explanation of the neutron lifetime anomaly. Additional constraints and expected limits (dashed) as per~\cite{McKeen:2020zni} have been included. This figure has been taken from~\cite{Berryman:2023rmh}.
    }
\end{figure}

In Figure~\ref{fig:narrow}, we focus on the particular range of masses pertinent to the neutron lifetime anomaly, thus reporting both $\epsilon_{n\chi}$ and ${\rm B}(n\to\chi \gamma)$. It is apparent that terrestrial searches for baryon number-violating processes in underground detectors give the most stringent limits on the possibility of $n\to \chi\gamma$, yet the windows on the $\chi$ mass open in such searches did not overlap with the possible mass window for the neutron lifetime anomaly---the nuclear stability constraint we have noted yielded the lowest possible $\chi$ mass, reported in Equation~(\ref{eq:chimasswindow}), and~this precludes that possibility. Nevertheless, we include the KamLAND~\cite{KamLAND:2005pen} and Super-Kamiokande~\cite{Super-Kamiokande:2015pys} limits, as~well as the limit on ${\rm B}(\Lambda\to \chi\gamma)$ from BESIII~\cite{BESIII:2021slv} and that from the SN1987A study of~\cite{Alonso-Alvarez:2021oaj} for context. Specifically, the~KamLAND constraints are relevant to our study of the decay $n\to \chi\gamma$; in these scenarios, the~prompt photon would not meet the correlation cuts, thus rendering it invisible. We observe that our neutron star limit is about $10^{18}$ times more severe than the accelerator limit and is similarly less sensitive to the BNV than the underground experiments. Additionally, we incorporated constraints arising from the influence of $\Lambda$ decay on the neutrino signal duration observed from SN1987A~\cite{Alonso-Alvarez:2021oaj}. These constraints have been derived from a supernova simulation involving an $18.6\, M_{\odot}$ progenitor leading to a $1.553\, M_{\odot}$ remnant~\cite{Garching}. Theoretically, these constraints could extend to lower $\chi$ masses as well. Notably, our derived limits significantly surpass the bounds established by the  supernova data~\cite{Alonso-Alvarez:2021oaj}.

The outcome of our analysis appears to exclude a new physics interpretation of the neutron lifetime anomaly, but~the context in which we performed this analysis is crucial: different choices within the dark sector could lead to alternate outcomes, as~explored in Section~\ref{sec:quasi-eq:NS}. In~our case, we have chosen a dark sector content and interaction as per Equation~(\ref{eq:chidark}) and~the accompanying discussion so that the SM dynamics of the neutron star determine the evolution of the star under a change in its baryon number---with the injected energy from the exotic neutron decay, particularly from $\chi$ production, removed through $\chi\chi \to \phi_B \phi_B$ decay, with~$\phi_B$ subsequently removed from the star. Our approach aligns with the depletion dominance route depicted in Figure~\ref{fig:NS_Dark_pathways}, thus resulting in a quasi-static evolution of the neutron star. However, in~alternative scenarios, the establishing limits would depend on varying pathways of dark decay, as~outlined in the same~figure.

As for dark cogenesis models, to~which our neutron dark decay studies connect, such as $B$ mesogenesis~\cite{Elor:2018twp}, we constrain the flavor structure of the quark--$\chi$ couplings through the limits we report~\cite{Berryman:2023rmh}---and we show our limits in our favored mass region according to the shaded band in Figure~\ref{fig:Limits_Combined}. 

% =======================================================
% =======================================================

% =======================================================
% =======================================================

% ========================================================
\section{Summary}
% ========================================================
There is an increasing recognition that the solution to the dark matter problem may involve an entire sector of dark particles with~dark dynamics. The~presence of such dynamics can be probed in terrestrial experiments that are sensitive to the determination of missing momentum~\cite{Berlin:2018bsc} in hadron decays,~and in this article, we have considered this broader possibility in the context of neutron stars. Particularly, we have noted the consequences from limits on the observation of energy loss in neutron stars, especially those that emerge from the timing of the period of binary pulsars on~baryon dark decays, specifically of neutrons and $\Lambda$ baryons. 

Neutron star observables have multiple points of contact with the problem of the neutron lifetime anomaly and~with the broader possibility of the simultaneous generation of dark matter and the BAU. We have noted models that would give rise to $n\to \chi \gamma$ or~to all dark decays, as~described in~\cite{Universe:Fornal,Universe:NS_Husain} in this volume, which are limited by the need to form neutron stars of sufficient mass. An~unwanted softening of the neutron EoS through such neutron decay effects can be mitigated through the possibility of $\chi$ self-interactions or~by reducing the neutron dark decay rate. We have studied the possibility of using dark sector interactions to eradicate the $\chi$ particles produced via dark decays through $\chi\chi$ annihilation to light, which are dark sector particles that can escape from the star. Here, we have studied the limit in which the dark decays proceed slowly with respect to the usual Urca processes that operate in the neutron star, thereby enforcing chemical and thermal equilibrium. Thus, in this quasi-static framework we are able to study the consequences of the observed energy loss constraints from pulsar timing studies on dark decays in the neutron stars. The~constraints we have found thus far are severe, but~broader possibilities remain---and they beckon. Certainly, there are models of the neutron lifetime anomaly that are not constrained by our concerns. Here, we note the model that yields $n\to \chi\chi\chi$~\cite{Strumia:2021ybk}. That model does not soften the EoS unduely, nor does it give rise to substantial energy loss effects that could be constrained through our pulsar timing studies. Thus, the possibility of a new physics solution to the neutron lifetime anomaly remains, with~the future neutron decay correlation and lifetime studies offering the possibility of setting the controversy to rest, though~effects in neutron stars, as well as in isolated ones, through observed heating or spin effects may finally serve to identify the effects of dark sectors in a positive~way. 

%%%%%%%%%%%%%%%%%%%%%%%%%%%%%%%%%%%%%%%%%%
\authorcontributions{All authors contributed equally to this~work. All authors have read and agreed to the published version of the manuscript.}

\funding{This research was supported by the U.S. Department of Energy Office of Nuclear Physics under contract~DE-FG02-96ER40989.}

\dataavailability{No new data were generated in the production of this article.} 

\acknowledgments{We thank Jeffrey M. Berryman for his past collaboration on many of the topics in this~overview.}

\conflictsofinterest{The authors declare no conflicts of~interest.} 

%%%%%%%%%%%%%%%%%%%%%%%%%%%%%%%%%%%%%%%%%%
\begin{adjustwidth}{-\extralength}{0cm}

\reftitle{References}

%=====================================
% References:
%=====================================

\PublishersNote{}
\end{adjustwidth}
\end{document}